\documentclass[pdftex,twocolumn,epjc3_preprint,runningheads]{svjour3}

\usepackage[T1]{fontenc}
\usepackage{lmodern}
\usepackage{calc}
\usepackage{slashed}
\usepackage{graphicx}
\usepackage{booktabs}
\usepackage{textcomp}
\usepackage{xspace}
\usepackage{relsize}
\usepackage{amssymb}
\usepackage{amsmath}
\usepackage{listings}
\usepackage{microtype}
\usepackage{multirow}
\usepackage{tabularx}
\usepackage{array}
\usepackage{placeins}
\usepackage{cuted}
\usepackage{soul} 
\usepackage{fixltx2e}
\usepackage[numbers,sort&compress]{natbib}
\usepackage[labelfont=bf,font=small]{caption}
\usepackage[skip=-2pt]{subcaption}
\usepackage[colorlinks,citecolor=blue,urlcolor=blue,linkcolor=blue,breaklinks=breakall]{hyperref}
\usepackage{breakurl}
\usepackage[dvipsnames]{xcolor}
\usepackage[clockwise,figuresright]{rotating}
\usepackage{siunitx}
\usepackage{tikz}
\usepackage{etoolbox}
\usepackage[dvipsnames]{xcolor}
\AfterEndEnvironment{strip}{\leavevmode}
\usepackage{textcomp}
\usepackage{makecell}
\usepackage{wrapfig}

\makeatletter
\newcommand{\preprintnumber}[1]{\gdef\@preprintnumber{\begin{flushright}{#1}\end{flushright}}}
\g@addto@macro\bfseries{\boldmath}
\makeatother

\makeatletter
\let\orgdescriptionlabel\descriptionlabel
\renewcommand*{\descriptionlabel}[1]{%
  \let\orglabel\label
  \let\label\@gobble
  \phantomsection
  \protected@edef\@currentlabel{#1}%
  \let\label\orglabel
  \orgdescriptionlabel{#1}%
}
\makeatother

\journalname{Eur. Phys. J. C}
\smartqed
\newcommand{\be}{\begin{equation}}
\newcommand{\ee}{\end{equation}}
\newcommand{\beq}{\begin{equation}}
\newcommand{\eeq}{\end{equation}}
\newcommand{\bea}{\begin{eqnarray}}
\newcommand{\eea}{\end{eqnarray}}
\newcommand{\beaa}{\begin{eqnarray*}}
\newcommand{\eeaa}{\end{eqnarray*}}
\newcommand{\ba}{\begin{array}}
\newcommand{\ea}{\end{array}}
\newcommand{\bi}{\begin{itemize}}
\newcommand{\ei}{\end{itemize}}
\newcommand{\ben}{\begin{enumerate}}
\newcommand{\een}{\end{enumerate}}

\newcommand{\ds}{\textsf{DarkSUSY}\xspace}

\newcommand{\db}{\textsf{DRAKE}\xspace}
\newcommand\Mma{\textsf{Mathematica}\xspace}
\newcommand\WE{\textsf{Wolfram Engine}\xspace}
\newcommand\WL{\textsf{Wolfram Language}\xspace}

\newcommand*{\courier}{\fontfamily{pcr}\selectfont}
\usepackage{changes}
\definechangesauthor[name={AH}, color=orange]{AH}

\begin{document}


\title{
{\color{Bittersweet}D}ark matter {\color{Bittersweet}R}elic {\color{Bittersweet}A}bundance beyond {\color{Bittersweet}K}inetic {\color{Bittersweet}E}quilibrium}

\author
{
Tobias Binder\thanksref{inst:a,e1}  \and
Torsten Bringmann\thanksref{inst:b,e2} \and
Michael Gustafsson\thanksref{inst:c,e3} \and
Andrzej~Hryczuk \thanksref{inst:d,e4} 
}

\institute{%
Kavli IPMU (WPI), UTIAS, The University of Tokyo, Kashiwa, Chiba 277-8583, Japan \label{inst:a} \and
Department of Physics, University of Oslo, Box 1048 Blindern, NO-0316 Oslo, Norway \label{inst:b}\and  
Institute for Theoretical Physics, Georg-August University G\"ottingen, Friedrich-Hund-Platz 1, D-37077 G\"ottingen, Germany \label{inst:c}
\and
National Centre for Nuclear Research, Pasteura 7, 02-093 Warsaw, Poland
\label{inst:d}}

\thankstext{e1}{tobias.binder@ipmu.jp}
\thankstext{e2}{torsten.bringmann@fys.uio.no}
\thankstext{e3}{michael.gustafsson@theorie.physik.uni-goettingen.de}
\thankstext{e4}{andrzej.hryczuk@ncbj.gov.pl}

\titlerunning{(running title)}
\authorrunning{(running author list)}

\date{ }

\maketitle

\begin{abstract}
We introduce \db, a numerical precision tool for predicting the dark matter relic abundance 
also in situations where the standard
assumption of kinetic equilibrium during the freeze-out process may not be satisfied.
\db comes with a set of three dedicated Boltzmann 
equation solvers that implement, respectively, the traditionally adopted
equation for the dark matter number density, fluid-like equations that couple the evolution of 
number density and velocity dispersion, and a full numerical evolution of the phase-space 
distribution. We review the general motivation for these approaches and, for illustration, highlight three 
concrete classes of models where kinetic and chemical decoupling are intertwined in a way that 
quantitatively impacts the relic density:  
{\it i)} dark matter annihilation via a narrow resonance, {\it ii)} Sommerfeld-enhanced
annihilation and {\it iii)} 
`forbidden' annihilation to final states that are kinematically inaccessible at threshold.
We discuss all these  cases in some detail, demonstrating that the commonly 
adopted, traditional treatment can result in an estimate of the relic density that is wrong by up to an order of magnitude. The public release of \db, along with several examples of how to calculate the relic density in concrete models, is provided at \href{https://drake.hepforge.org/}{drake.hepforge.org}

\newpage
\end{abstract}

\tableofcontents

\newpage
\section{Introduction} 

The most often studied scenario to explain the observed cosmological dark matter (DM) 
abundance~\cite{Aghanim:2018eyx} 
considers new elementary particles that have been thermally produced in the early 
universe. This freeze-out scenario~\cite{Lee:1977ua} provides an intriguing solution to the DM puzzle
not only for classical Weakly Interacting Massive Particle (WIMP) candidates~\cite{Lee:1977ua,Ellis:1983ew,Jungman:1995df,Hooper:2007qk,Arcadi:2017kky},
with masses at the electroweak scale, but also for much lighter DM particles (which is sometimes referred 
to as `WIMP-less miracle'~\cite{Feng:2008ya}). Various numerical codes are available to provide
precision calculations of the DM relic density in these models, including public tools 
like \ds~\cite{Bringmann:2018lay}, 
{\textsf{MadDM}\xspace}~\cite{Ambrogi:2018jqj}, {\textsf{micrOMEGAs}\xspace}~\cite{Belanger:2006is} 
and {\textsf{SuperISOrelic}\xspace}~\cite{Arbey:2009gu}. In fact, precision calculations are necessary in 
order to match the percent-level observational accuracy. In global fits of the full underlying model 
parameter space, for example,
the relic density often provides one of the most relevant observables in terms of 
contributions to the total likelihood~\cite{Roszkowski:2014wqa,Bechtle:2015nua,Baer:2016ucr,Athron:2017qdc, Athron:2017yua, Athron:2017kgt,Bagnaschi:2017tru,Costa:2017gup,Bagnaschi:2019djj}.

All these codes currently rely on the standard approach~\cite{Gondolo:1990dk,Edsjo:1997bg} 
to calculate the relic density,\footnote{%
 As of version 6.2.5, along with the release of \db, \ds additionally provides numerical
routines to solve the coupled system of Boltzmann equations described in Section \ref{sec:fluid}.
} 
which can be re-cast in the
form of a single Boltzmann equation for the DM number density even when including sharp resonances, 
thresholds or co-annihilations (all of which were initially considered complications~\cite{Griest:1990kh}).
One of the main underlying assumptions for this formulation is that DM (along with potential 
co-annihilating particles) is kept in kinetic equilibrium with the standard model (SM) heat bath during the 
entire chemical decoupling process. There exist however various well-motivated scenarios where this 
assumption is not satisfied, at least not for all relevant parts of the parameter space, and where the 
standard approach is thus not directly applicable. \mbox{A possible} solution that has received increased attention 
is to derive coupled equations for the number density and higher moments of the DM phase-space 
distribution, making use  of the Boltzmann 
hierarchy \cite{vandenAarssen:2012ag,Duch:2017nbe,Kamada:2017gfc,Berlin:2018tvf,Kamada:2018hte,Abe:2020obo}. 
More recently, there have also been attempts to solve the full Boltzmann equation directly at the 
phase-space level~\cite{Binder:2017rgn,Brummer:2019inq,Ala-Mattinen:2019mpa}.

Here we introduce a new public code, \href{drake.hepforge.org}{\db}, that is written in 
\textsf{Wolfram Language}\footnote{%
\db can thus be directly used within \href{https://www.wolfram.com/mathematica/}{\textsf{Mathematica}}, 
but its implementation also allows, without loosing any functionality, for a script usage with the free 
\href{https://www.wolfram.com/engine/}{\textsf{Wolfram Engine}}. 
} 
and implements both of these more general approaches for the 
case of annihilating DM. \db thus complements
existing codes to calculate the relic density also 
in situations where the underlying assumptions of the traditional 
approach are not satisfied. Additionally, it allows in fact to examine 
the validity of these assumptions explicitly.
As an application, we also study in some detail three concrete physics scenarios where kinetic decoupling 
can interfere  with the freeze-out process: {\it i)} $s$-channel resonances, {\it ii)} Sommerfeld 
enhancement and {\it iii)} sub-threshold annihilation. In all these cases, we contrast the results of classical 
relic density calculations with the more accurate results obtained by \db.

This article is organized as follows. We start in Section \ref{sec:scope} by introducing the relevant 
Boltzmann equations to describe the interaction of DM particles with the thermal bath. In 
Section \ref{sec:code}, the structure of the code is introduced, including the main implemented algorithms 
that are used to solve these Boltzmann equations. We then discuss in some detail the results of relic 
density calculations in concrete physics applications in 
Section~\ref{sec:exampels}, before concluding in Section \ref{sec:summary}. In two Appendices we 
provide a quick-start guide for using \db (App.~\ref{app:quick_start})
and discuss examples of elastic scattering operators that can be treated beyond the commonly 
adopted Fokker-Planck approximation (App.~\ref{app:full_cel}).

\section{Scope}
\label{sec:scope}

\subsection{Full Boltzmann equation}

We will consider situations where DM interactions with the SM heat bath, through elastic
scattering and annihilation processes, are initially strong enough to establish both chemical
and kinetic equilibrium. As the universe expands, the DM particles (denoted by $\chi$) 
fall out of equilibrium and eventually establish the present relic abundance.
The evolution of the DM phase-space density $f_\chi(t,p)$ during this process
is governed by what we will refer to as the full Boltzmann equation (\textbf{fBE}):
\be
  \label{diff_boltzmann}
  E\left(\partial_t-Hp\partial_p\right)f_\chi=C_{\rm ann}[f_\chi] + C_{\rm el}[f_\chi]\,,
\ee
where
\bea
  \label{Candd_def}
  C_\mathrm{ann}&=&\frac{1}{2g_\chi}\int\frac{d^3\tilde p}{(2\pi)^32\tilde E}\int\frac{d^3k}{(2\pi)^32\omega}\int\frac{d^3\tilde k}{(2\pi)^32\tilde \omega}\\
&&\times(2\pi)^4\delta^{(4)}(\tilde p+p-\tilde k-k)\nonumber\\
&&\times\left[
\left|\mathcal{M}\right|^2_{\bar\chi\chi\leftarrow \bar f f}g(\omega)g(\tilde \omega)
-\left|\mathcal{M}\right|^2_{\bar\chi\chi\rightarrow \bar f f}f_\chi(E)f_\chi(\tilde E)
\right]\nonumber
\eea
describes the effect of two-body annihilations, 
and the collision term for elastic scattering processes is given by
\bea
  \label{Celd_ef}
  C_\mathrm{el}&=&\frac{1}{2g_\chi}\int\frac{d^3k}{(2\pi)^32\omega}\int\frac{d^3\tilde k}{(2\pi)^32\tilde \omega}\int\frac{d^3\tilde p}{(2\pi)^32\tilde E}\\
  &&\times(2\pi)^4\delta^{(4)}(\tilde p+\tilde k-p-k) {\left|\mathcal{M}\right|}^2_{\chi f\leftrightarrow\chi f}\nonumber\\
  &&\times\left[\left(1\mp g^\pm(\omega)\right)\, g^\pm(\tilde\omega)f_\chi({\tilde E})-
  (\omega\leftrightarrow\tilde\omega, {E}\leftrightarrow{\tilde E})\right]\,.\nonumber
\eea
In the above expressions, $H=\dot a/a$ is the Hubble parameter, $a$ the scale factor, and
we have assumed a 
Friedman-Robertson-Walker universe, such that $f_\chi$ only depends on the
absolute value of the DM momentum, $p=|\mathbf{p}|$.
Furthermore, both collision terms and the squared amplitudes ${\left|\mathcal{M}\right|}^2$ for the 
respective process are implicitly summed over all heat bath particles $f$, and final {\it and} initial state 
internal degrees of freedom, respectively. The phase-space distribution of the heat bath particles is given
by the usual $g^\pm(\omega)=1/\left[\exp(\omega/T)\pm1 \right]$. Since we assume that DM is 
non-relativistic around freeze-out, we have neglected Bose enhancement and Pauli blocking factors for 
$f_\chi$ (which implies that these factors are also negligible for the heat bath particles in 
$C_\mathrm{ann}$ due to energy conservation). For further details about Eq.~(\ref{diff_boltzmann}), 
see Refs.~\cite{Bernstein:1988bw,Kolb:1990vq}.

\subsection{Evaluating the collision terms}
\label{sec_collision}
In practice, the greatest obstacle in solving Eq.~(\ref{diff_boltzmann}) to sufficient precision is often the 
evaluation of the collision integrals on the right-hand side. For the annihilation term, this is somewhat less 
critical since -- under the generic assumptions of $CP$ invariance and $f_\chi\ll1$ 
-- the phase-space integrals can always be reduced to only one remaining
angular integration, and one in energy~\cite{Gondolo:1990dk}:
\bea
  \label{Cann_simp}
 C_\mathrm{ann}&=& g_{\chi} E\int\frac{d^3\tilde p}{(2\pi)^3} \,v_{\rm M} \sigma_{\bar\chi\chi\rightarrow \bar f f}\nonumber\\
  &&\times\left[
f_{\chi,{\rm eq}}(E)f_{\chi, {\rm eq}}(\tilde E)-f_\chi(E)f_\chi(\tilde E)
\right]
\,, 
\eea
where the non-relativistic DM particles in equilibrium follow a Maxwell-Boltzmann 
distribution, $f_{\chi,{\rm eq}}(E)=e^{-E/T}$. Further,
$\sigma$ is the annihilation cross-section for the process $ \bar\chi\chi\rightarrow \bar f f $, 
and 
$v_{\rm M}\equiv [s(s-4m_\chi^2)]^{1/2}/({2 E \tilde E})$ is the M\o ller velocity;
in the rest frame of either of the DM particles, it equals the velocity of the other particle, 
$v_{\rm M}=v_{\rm lab}\equiv[s(s-4m_\chi^2)]^{1/2}/{(s-2m_\chi^2)}$. 
For later reference, let us here also introduce the angular ($\tilde{\Omega}$) averaged 
quantity $\langle \sigma v \rangle_{\theta} \equiv \frac{1}{2} \int \text{d} \cos \theta \,v_{\rm M} \sigma_{\bar\chi\chi\rightarrow \bar f f }$ 
that is used in our numerical code and depends only on the magnitude of the three-momenta, 
$p$ and $\tilde{p}$. 

The elastic scattering term also simplifies for highly non-relativistic DM.  For $m_f\ll m_\chi$, 
specifically, the typical momentum transfer per collision is then much smaller than the average DM 
momentum in equilibrium. Expanding $C_\mathrm{el}$ up to second order in the momentum transfer 
results in a simple differential operator of 
Fokker-Planck type~\cite{Bertschinger:2006nq, Bringmann:2006mu, Bringmann:2009vf}, which can
be used to describe kinetic decoupling taking place much later than chemical decoupling (see, e.g., 
Ref.~\cite{Bringmann:2016ilk}). To improve 
the description of early kinetic decoupling, we keep in \db some of the leading relativistic 
corrections~\cite{Binder:2016pnr,Binder:2017rgn}, resulting in the Fokker-Planck operator~(FP): 
\begin{align}
C_\mathrm{el} & \simeq C_\mathrm{FP} \label{eq:fokkerplanckrel} \\ &=
\frac{E}{2} \gamma(T)
{\Bigg [}
T E\partial_p^2 \!+ \left(2 T  \frac{E}{p} \!+\! p \!+\! T\frac{p}{E}\right) \partial_p + 3
{\Bigg ]}f_{\chi}\,.\nonumber
\end{align}
It is important to note that $C_\mathrm{FP}[f_{\chi}]=0$ for a relativistic Maxwellian shape 
$f_{\chi} \propto e^{-E/T}$, which is consistent with the stationary solution of the relativistic annihilation 
term in Eq.~(\ref{Cann_simp}) (see appendix in Ref.~\cite{Binder:2017rgn} for more details).
Furthermore, it is worth mentioning that the above operator can be written as a total momentum 
divergence 
and hence manifestly conserves the DM particle number. 
The momentum transfer rate $\gamma(T)$ introduced above is
the same as in the highly non-relativistic version of Eq.~(\ref{eq:fokkerplanckrel}), and given by (see also 
Refs.~\cite{Kasahara:2009th, Gondolo:2012vh}) 
\begin{align}
  \label{cTdef}
 \gamma= \! \frac{1}{3 g_{\chi} m_{\chi} T} \! \int \! \frac{\text{d}^3 k}{(2\pi)^3} g^{\pm}(\omega)\left[1\!\mp\! g^{\pm}(\omega)\right] \! \! \! \int\limits^0_{-4 k_\mathrm{cm}^2} \! \! \!  \text{d}t (-t) \frac{\text{d}\sigma}{\text{d}t} v\,,
\end{align}
where ${k}_\mathrm{cm}^2  \equiv m_\chi^2{k}^2/(m_\chi^2+2\omega m_\chi+m_f^2)$ and the scattering 
amplitude entering the differential cross-section, 
$({\text{d}\sigma}/{\text{d}t}) v \equiv |\mathcal{M}|^2_{\chi f\leftrightarrow\chi f}/(64 \pi {k} \omega m_\chi^2)$, 
is evaluated at $s\simeq m_\chi^2+2\omega m_\chi+m_f^2$. 

We will encounter one concrete example in this work, in Section \ref{sec:sub_threshold}, where the mass 
of the scattering partner is comparable to the DM mass, such that the momentum transfer can no longer 
be assumed to be small. In this case, we have to resort to the full scattering term in Eq.~(\ref{Celd_ef}), 
which we re-write in a form that is more suitable for numerical integration, and which formally resembles 
the annihilation case, Eq.~(\ref{Cann_simp}):\footnote{%
To arrive at this form we use
$g^{\pm}(\tilde{\omega})\left[1 \mp g^{\pm}(\omega)\right] = e^{-(E-\tilde{E})/T}$ $g^{\pm}(\omega)\left[1 \mp g^{\pm}(\tilde{\omega})\right] $, 
as  implied by energy conservation.
}
\begin{align}
C_\mathrm{el}= & E\int \frac{\text{d}^3\tilde p}{(2\pi)^3}   W \label{eq:Cel}\\
&\times\left[e^{-(E-\tilde{E})/(2T)} f_{\chi}(\tilde{E})- e^{(E-\tilde{E})/(2T)} f_{\chi}(E)\right]. \nonumber
\end{align}
Here, the quantity $W$ is defined as
\begin{align}
W \equiv &  \frac{e^{-(E-\tilde{E})/(2T)}}{4 g_{\chi}  E \tilde{E}}  \int\!\! \frac{\text{d}^3{k}}{(2 \pi)^3 2 \omega} \frac{\text{d}^3\tilde{{k}}}{(2 \pi)^3 2 \tilde{\omega}}  g^{\pm}(\omega)  \left[1\!\mp\! g^{\pm}(\tilde{\omega}) \right]  \nonumber \\
&  \times (2\pi)^4\delta^4(p+k-\tilde{p}-\tilde{k})|\mathcal{M}|^2_{\chi f\leftrightarrow\chi f}
\label{eq:Wmain}
\end{align}
and has the same physical dimension as a cross-section. For our numerical codes it is convenient to 
factorise out its angular average 
$ \langle W \rangle_{\tilde{\Omega}} \equiv \frac{1}{4 \pi}\int \text{d} \tilde{\Omega}\, W$, depending only 
on $T$, $p$ and $\tilde{p}$. This quantity is in general challenging to compute due to the 
multidimensional integrals. In Appendix~\ref{app:full_cel} we summarize, however, some concrete cases 
(including the one relevant for the discussion in Section \ref{sec:sub_threshold}) where the amplitude 
$\mathcal{M}$ takes a form allowing this general expression for $C_\mathrm{el}[f]$ to be analytically 
reduced to a one-dimensional integral.

\subsection{Fluid equations}
\label{sec:fluid}

An alternative to the numerically challenging task of directly computing the evolution of the phase-space 
density $f_\chi$  consists in restricting the discussion to its first moments.  Instead of the full 
Boltzmann Eq.~(\ref{diff_boltzmann}) one thus only considers the corresponding momentum moments of 
this equation, thereby implementing a `hydrodynamical' approach to the problem that essentially results in 
a set of fluid equations. Just as in the case of hydrodynamics, 
however, additional assumptions are needed to close the Boltzmann hierarchy at any given level. 

The simplest case, and in fact the traditional way to handle relic density calculations, is to only consider 
the lowest moment of $f_\chi$, namely the DM number density 
$n\equiv g_\chi \int d^3p\,f_\chi /(2\pi)^{3}$. The additional requirement to close the resulting equation 
for $n$ is typically met by assuming that kinetic equilibrium is maintained until the chemical decoupling 
process is completed.  For non-relativistic DM, this implies that the DM phase-space distribution is of 
the form $f_\chi= \exp[(\mu-E)/T]$
, which fixes $f_\chi$ up to a function of $T$ 
(here parameterized as the DM chemical potential $\mu$, in this case related to the number density as $\mu/T=\ln[n/n^{\text{eq}}]$).

Inserting this form into Eq.~(\ref{diff_boltzmann}), and integrating over the DM momenta $\mathbf{p}$,
then results in what we will refer to as the traditional 
number density equation (\textbf{nBE}):
\begin{align}
\label{eq:nBE}
\dot{n} + 3Hn = -\langle \sigma v \rangle_T \left[ n^2 - n_{\text{eq}}^2(T) \right]\,,
\end{align}
where $n_{\text{eq}}$ refers to the DM density in chemical equilibrium, i.e.~for $\mu=0$. 
The thermal average $\langle...\rangle_T$ of the velocity-weighted annihilation cross-section can be 
stated in terms of a single integral over the center-of-mass energy, as explicitly given in Eq.~(3.8) of Ref.~\cite{Gondolo:1990dk}.

In situations where kinetic decoupling interferes with the chemical decoupling process, the main 
assumption leading to Eq.~(\ref{eq:nBE}) is clearly too 
restrictive. This implies that one needs to move (at least) one level up in the Boltzmann hierarchy to 
(approximately) describe such situations, and thus to leave the second moment of $f_\chi$ as a dynamical
degree of freedom. 
A convenient parameterization for this is the DM velocity dispersion or `temperature', 
$T_\chi\equiv g_\chi/(3n_\chi)\int d^3p\,(2\pi)^{-3} (p^2/E) f_\chi$,
and one way of closing the Boltzmann hierarchy at this level is to assume, in analogy to the
case discussed above, $f_\chi=
\exp[(\mu-E)/T_\chi]$ (with $\mu/T_\chi=\ln[n/n_{\rm eq}(T_\chi)]$; 
see Ref.~\cite{Binder:2017rgn}  for a more detailed discussion).
Using Eq.~(\ref{eq:fokkerplanckrel}), and assuming entropy conservation, 
this results in what we will refer to as the coupled Boltzmann equations (\textbf{cBE}):
\bea
\frac{Y'}{Y} &=& \frac{s Y}{x \tilde H}\left[
\frac{Y_{\rm eq}^2}{Y^2} \left\langle \sigma v\right\rangle_T- \left\langle \sigma v\right\rangle_{T_\chi}
\right]\,, \label{Yfinalfinal}\\
\frac{y'}{y} &=&   \frac{1}{x\tilde H}\langle C_{\text{el}} \rangle_2
+\frac{sY}{x\tilde H}\left[
\left\langle \sigma v\right\rangle_{T_\chi}-\left\langle \sigma v\right\rangle_{2,T_\chi}
\right] \label{yfinalfinal}\\ 
&&+\frac{sY}{x\tilde H}\frac{Y_{\rm eq}^2}{Y^2}\left[
\frac{y_{{\rm eq}}}{y}\left\langle \sigma v\right\rangle_{2,T}\!-\!\left\langle \sigma v\right\rangle_T
\right]
+2(1-w)\frac{H}{x\tilde H}\,, \nonumber
\eea
where $w(T_\chi)\equiv 1-{\langle p^4/E^3 \rangle_{T_\chi}}/({6T_\chi})$ parameterizes the
deviation from the highly non-relativistic case ($w=1$).
In order to arrive at this compact form, we have followed the standard convention of introducing dimensionless variables $Y(x)\equiv n/s$,  $Y_{\rm eq}(x)\equiv n_{\rm eq}(T)/s$,
$y(x)\equiv m_\chi T_\chi s^{-2/3}$ and $x\equiv m_\chi/T$, where $s$ is the 
entropy density.\footnote{%
The value of $Y$ today, $Y_0\equiv Y(x\to \infty)$, relates to the present abundance of $\chi$ via 
$\Omega_\chi h^2= 2.755\times10^{10}\left(\frac{m_\chi}{100\,{\rm GeV}}\right) \left(\frac{T_{\rm CMB}}{2.726\,{\rm K}}\right)^3Y_0$~\cite{Gondolo:1990dk}. 
For Dirac DM particles, the total DM abundance is thus given by 
$\Omega_{\rm DM}=2\Omega_\chi$.
}
The thermal average $\langle \sigma v\rangle_{2,T}$ is a variant of 
the commonly used thermal average $\langle \sigma v\rangle_T$, and is explicitly stated in 
Ref.~\cite{Binder:2017rgn}. We also introduced
\begin{align}
\langle C_{\text{el}} \rangle_2 \equiv \frac{g_\chi}{3 n T_{\chi}} \int \frac{\text{d}^3 p}{(2\pi)^3} \frac{p^2}{E^2} C_{\text{el}}\;,
\end{align}
which in the Fokker-Planck approximation of Eq.~(\ref{eq:fokkerplanckrel}) simplifies to 
$\langle C_{\text{el}} \rangle_2 \rightarrow \gamma(T) w(T_{\chi}) \left[(y_{{\rm eq}}/y) -1\right]$.
Finally, $\tilde H\equiv H/\left[1+ (1/3)d(\log g^s_{\rm eff})/d(\log T)\right]$, 
with $g^s_{\rm eff}$ being the entropy degrees of freedom of the 
background plasma. The first of these equations, Eq.~(\ref{Yfinalfinal}), is the analogue of the traditional number density equation; in the limit $T_{\chi}\rightarrow T$ (kinetic equilibrium) it reduces as expected exactly to Eq.~(\ref{eq:nBE}) 
expressed in dimensionless variables. 

Let us close this section by briefly mentioning the potential effect of DM 
self-interactions~\cite{Spergel:1999mh,Tulin:2017ara} on  
our discussion. In principle, this would be described by an additional collision term $C_{\rm self}$ 
on the right-hand side of Eq.~(\ref{diff_boltzmann}), but bringing such a term into a form that is 
numerically tractable is challenging. In light of this situation it is interesting to note that the two 
beyond-\textbf{nBE} approaches implemented in \db can be regarded as an effective way of bracketing 
the impact of such model-dependent 
DM self-interactions: our implementation of \textbf{fBE} provides the correct description of how the DM 
phase-space density evolves {\it if}  the effect of DM self-interactions is negligible, while the Eqs.~(\ref{Yfinalfinal},\ref{yfinalfinal}) that \textbf{cBE} is based on become {\it exact} under the assumption that 
DM self-interactions are maximally efficient and hence force the DM distribution to be of the form  
$f_\chi=\exp[(\mu-E)/T_\chi]$.

\section{Code design} 
\label{sec:code}
\db is a package of routines written in \textsf{Wolfram Language} that allows to numerically compute the 
 DM relic abundance, $\Omega_{\chi}h^2$, in the \textbf{nBE}, \textbf{cBE} and \textbf{fBE} 
 approaches introduced above. In Section~\ref{sec:drake}, we provide an overview of
the code's modular structure and how to use it in practice to obtain $\Omega_{\chi}h^2$ and other related 
quantities for a given DM particle model.
The internal algorithm implemented 
for the time ($x$) integration of the three Boltzmann equations is presented in 
Section~\ref{sec:impl}, and in Section~\ref{sec:test} we summarize various validity checks 
of the code that we have performed.

\subsection{Overview} \label{sec:drake}

The \db package provides routines to \emph{solve} (time integrate) the 
individual Boltzmann equations, as well as a set of \emph{preparatory} routines that compute quantities required  by the solvers, e.g., averages of the annihilation cross-section. This design structure allows to perform relic abundance 
computations in a flexible manner, to reuse certain output quantities from preparing routines in multiple 
Boltzmann solvers, and reduce the required model input to fairly simple expressions.

To make the code design 
concrete, let us start with the three Boltzmann \emph{solver} routines {\courier nBE}, {\courier cBE} and 
{\courier fBE} listed in Table~\ref{tab:aroutines}. As explained in Section~\ref{sec:impl}, these are adaptive 
and implicit ordinary differential equation solvers, where the \emph{partial} differential 
Eq.~(\ref{diff_boltzmann}) for the {\courier fBE} solver has been mapped to a set of coupled ordinary 
differential equations by the so-called method of lines (see, e.g., Ref.~\cite{MOL_book}).
Each \emph{solver} calculates
the relic abundance $\Omega_\chi h^2$ and the 
full $Y(x)$ evolution. {\courier cBE} and  {\courier fBE} also give the $y(x)$ evolution, and 
the latter additionally provides information on the full phase-space density $f_\chi(x,p)$.

The annihilation cross-section averages (third column) needed in the  Boltzmann equations are computed 
by the \emph{preparatory} routines {\courier PrepANN}, {\courier PrepANN2}, and {\courier 
PrepANNtheta}, as listed in Table~\ref{tab:broutines}. The only input required from the user
is the DM annihilation cross-section 
in the form of $\sigma v_{\text{lab}}$ as a function of the Mandelstam variable $s$.
The routines then adopt to adjustable accuracy settings and the DM model at hand, in order to 
generate accurate and densely tabulated outputs for the resulting interpolation functions.

For the {\courier cBE} and {\courier fBE} routines also the scattering operator $C_{\text{el}}$ in 
Eq.~\eqref{Celd_ef} needs to be evaluated. If the Fokker-Planck approximation in 
Eq.~\eqref{eq:fokkerplanckrel} is chosen for the scattering operator, then 
the momentum transfer rate $\gamma(x)$ in Eq.~\eqref{cTdef} --- or just the model-dependent part containing the 
Mandelstam $t$-average of the scattering amplitude $|\mathcal{M}|^2_{\chi f\leftrightarrow\chi f}$ --- 
should be provided by the user. The preparatory routine {\courier PrepSCATT}
then adaptively tabulates $\gamma(x)$ to provide an interpolating function for fast evaluation. 
However, we stress that the {\courier cBE} and {\courier fBE} routines are not limited to the use of the 
Fokker-Planck approximation for the scattering collision term $C_{\text{el}}$.
As explained in Appendix~\ref{app:full_cel}, the full $C_{\text{el}}$ can be used if the quantity 
$\langle W \rangle_{\tilde\Omega}$ from Eq.~\eqref{eq:Wmain} is provided (which in practice 
amounts to providing $\langle C_{\text{el}} \rangle_2$ and $\hat{\mathbf{C}}_{\text{el}}$ 
as defined in \ref{app:implementation}
for {\courier cBE} and {\courier fBE}, respectively).

\begin{table}[t]
\begin{center}
\begin{tabular}{l||l|c|c} 
\toprule 

   Routine & ~~~solves & \multicolumn{1}{c}{}  & \hspace{-2cm}based on\\ \midrule
    ~~{\courier nBE} & Eq.~(\ref{eq:nBE}) & $\langle \sigma v\rangle$ & - \\ \midrule 
    ~~{\courier cBE} & Eq.~(\ref{Yfinalfinal},\ref{yfinalfinal}) & $\langle \sigma v\rangle$, $\langle \sigma v\rangle_2$ & $\gamma$ or $ \langle W \rangle_{\Omega}$ \\  \midrule 
    ~~{\courier fBE} & Eq.~(\ref{diff_boltzmann}) & $ \langle \sigma v\rangle_{\theta}$ & $\gamma$ or $ \langle W \rangle_{\Omega}$ \\ \bottomrule
\end{tabular}
\end{center}
\caption{The names of the three routines \emph{solving}, respectively, the Boltzmann equation
referred to in the second column. The third and fourth column show the basic
quantities that these equations are based on. These routines are, respectively, located in 
the files \textbf{nBE.wl}, \textbf{cBE.wl} and \textbf{fBE.wl} (in the src/ directory).
}
\label{tab:aroutines}
\end{table}

\begin{table}
\begin{center}
\begin{tabular}{l||l|l} 
\toprule 
   ~Routine & computes & as needed for \\ \midrule
    {\courier PrepANN} & $\langle \sigma v\rangle(x)$    & {\courier nBE}, {\courier cBE} \\ \midrule

    {\courier PrepANN2}  & $\langle \sigma v\rangle_2(x)$ &{\courier cBE} \\ \midrule

    {\courier PrepANNtheta}  & $ \langle \sigma v\rangle_\theta(p,\tilde{p})$ & {\courier fBE} \\ \midrule
    
    {\courier PrepSCATT}  & $\gamma(x)$ & {\courier cBE}, {\courier fBE}     \\  \bottomrule
\end{tabular}
\end{center}
\caption{The \emph{preparatory} routines, computing averages of the velocity-weighted 
annihilation cross-section $\sigma v_{\text{lab}}$.
Output of these routine calls is stored in the session memory, such that, e.g., the quantity 
$\langle \sigma v\rangle$ can be used for both {\courier nBE} and {\courier cBE} routine calls. These 
routines are located in \textbf{rates.wl} (in the src/ directory).}
\label{tab:broutines}
\end{table}

The first code release also comes with five pre-implemented DM model setups. 
These are the Scalar Singlet DM model~\cite{Silveira:1985rk,McDonald:1993ex,Burgess:2000yq}
\footnote{Our Singlet Scalar model implementation is specified in Ref.~\cite{Binder:2017rgn}, with the annihilation 
cross-section updated to now follow what is currently adopted in \ds.
}
,
our three example scenarios discussed in Section \ref{sec:exampels}, 
and a WIMP-like toy model. 
Each of these comes with three files: a \emph{model} file (containing $\sigma v_{\text{lab}}(s)$ and either 
$\gamma(x)$ or the scattering operators $\langle C_{\text{el}} \rangle_2$ and $\hat{\mathbf{C}}_{\text{el}}$), 
a \emph{parameters} file  (with numerical values of the DM mass, couplings, internal d.o.f., etc.) and 
a \emph{settings} file (with various code options).

\begin{figure*}[t]
\begin{center}
\includegraphics[scale=0.27]{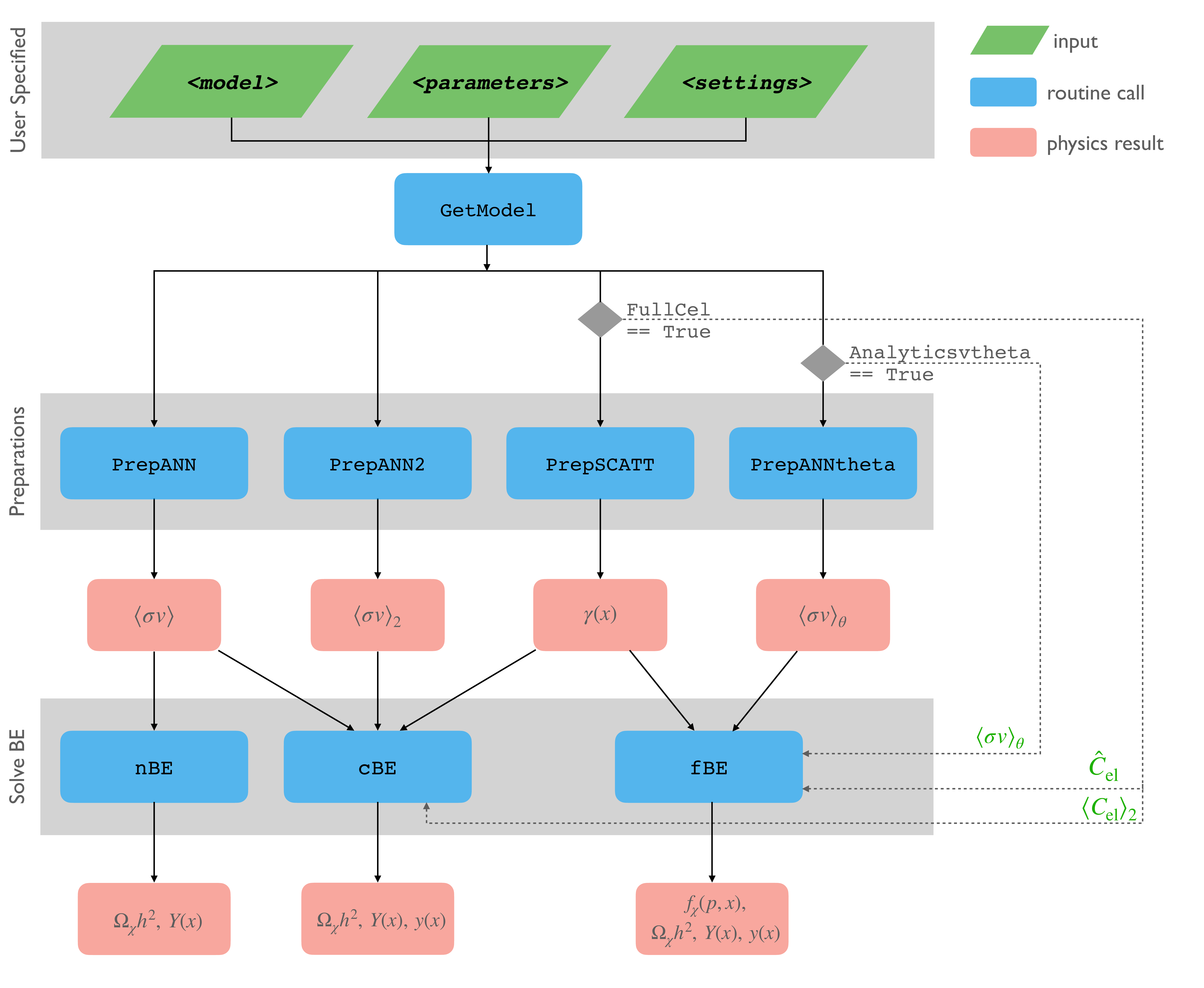}
\caption{Flowchart showing \db's main routines as called by the \textbf{main.wls} template script. 
The three user specified input \textbf{\textit{\courier <files>}} in the green boxes should contain the DM 
\emph{model} definition, numerical values on model \emph{parameters} and 
{option}
\emph{settings} 
for \db (see main text and 
Appendix~\ref{app:quick_start}).
The command {\courier GetModel} then loads the content from these \textbf{\textit{\courier <files>}} 
before the script continues to execute the required \emph{preparatory} routines. 
The preparatory routines {\courier PrepANNtheta} and {\courier PrepSCATT} will not be called if 
{the options} {\courier 
Analyticsvtheta} or {\courier FullCell} are set to {\courier True}, respectively; 
in that case the quantities $\langle \sigma v \rangle_\theta$ or, respectively, $\hat{\mathbf{C}}_{\text{el}}$ 
and  $\left \langle C_\text{el} \right \rangle_2$, need to be provided by (any of) the user defined 
\textbf{\textit{\courier <files>}} (as indicated by the green texts along the dashed lines). 
The main physics output of each preparatory routine is stated in the respective red box right below,
and the subsequent calls of the Boltzmann equation solvers {\courier nBE}, 
{\courier cBE} and {\courier fBE} depend on these physical quantities as indicated by their incoming arrows.
The  output from each Boltzmann solver, finally, is explicitly stated in the red boxes in the bottom row. 
} 
\label{fig:code}
\end{center}
\end{figure*}

The DM relic abundance in these, or any other user-defined, setups can be conveniently calculated, e.g.\ within 
a \Mma notebook, by 
sequentially calling the required \db routines.
A practical alternative is to use the customizable template script \textbf{main.wls} provided in the main directory.
As illustrated in Fig.~\ref{fig:code}, 
the calculation procedure consists of initializing  the \db package, then loading the DM model setup files, performing the required \emph{preparatory} 
calculations and finally executing the Boltzmann equation \emph{solvers}.

A dedicated quick start guide as well as a description of more features and details of \db are provided in  Appendix~\ref{app:quick_start}.

\subsection{Numerical implementation}
\label{sec:impl}
All three Boltzmann equations (\textbf{nBE}, \textbf{cBE} and \textbf{fBE}) are numerically time ($x$) integrated 
by essentially the same implicit adaptive algorithm. Once the phase-space density is discretized in 
momentum space as $f_{\chi}(t,p) \rightarrow \{f_0(t),...,f_N(t) \}$, in particular, all of them can be written 
as a coupled system of ordinary first order differential equations of the form
\begin{align}
\frac{\text{d}}{\text{d}x} \mathbf V(x) = \mathbf{F}(x, \mathbf V),
\end{align}
where 
$ \mathbf V(x) = \{V_0(x), ...,V_N(x) \}$ and $\mathbf{F} = \{ F_0(x, \mathbf V), ...,F_N(x, \mathbf V) \}$. 
For concreteness, $ \mathbf V(x) = Y(x)$ is one-dimensional for \textbf{nBE}, $ \mathbf V(x) = \{Y(x),y(x)\}$ 
for \textbf{cBE} as in Eqs.~(\ref{Yfinalfinal}) and (\ref{yfinalfinal}), and $ \mathbf V(x) = \{f_0(x),...f_N(x)\}$
for the momentum-discretized phase-space density in the \textbf{fBE} approach.
We follow standard practice for the numerical 
integration of these equations, see, e.g., Ref.~\cite{Nbook}.

Concretely, we choose the Adams-Moulton time discretization methods 
of order one and two, which are also known as the implicit Euler 
\begin{align}
& \mathbf V_{i} = \mathbf V_{i-1} + h \left[ \mathbf F(x_{i}, \mathbf V_{i})\right], \label{eq:Euler}
\end{align}
and implicit trapezoidal method
\begin{align}
 \mathbf V_{i} = \mathbf V_{i-1} + \frac{h}{2} \left[ \mathbf F(x_{i-1}, \mathbf V_{i-1})+ \mathbf F(x_{i}, \mathbf V_{i})\right], \label{eq:Trapezoidal}
\end{align}
where $ \mathbf V_i\equiv \mathbf V(x_i)$ and $x_{i}=x_{i-1}+h$. The adaptive step size $h$ is controlled through a 
local relative error of the two methods, 
$\text{err}_i \equiv \underset{j}{\text{Max}}\{\text{Abs}\left[( (\mathbf V_{i}^T)_j - (\mathbf V_{i}^E)_j )/( \mathbf V_{i}^T )_j \right]\}$
where $ \mathbf V_{i}^E$ and $ \mathbf V_{i}^T$ are the solutions of the implicit Euler Eq.~(\ref{eq:Euler}) 
and trapezoidal Eq.~(\ref{eq:Trapezoidal}), respectively. By default, the solution $ \mathbf V_{i}^T$ is 
accepted if $\text{err}_i < 10^{-3}$. 

For the single number density equation (\textbf{nBE}), the implicit equations (\ref{eq:Euler}) and 
(\ref{eq:Trapezoidal}) can respectively be solved analytically  
for $ \mathbf V_{i}^T$ and $ \mathbf V_{i}^E$, allowing for efficient 
time integration. In this one-dimensional case, thus, the implemented algorithm in our {\courier nBE} routine is equal to the one used in \ds.

In the case of \textbf{cBE} and \textbf{fBE}, the solution of
Eqs.~(\ref{eq:Euler},\ref{eq:Trapezoidal}) needs to be obtained numerically. A standard root finding 
algorithm is the Newton iteration method. Defining the iteration depth $n$ and introducing 
$\mathbf \Delta_{i}^{(n+1)}  \equiv  \mathbf V^{(n)}_{i} - \mathbf V^{(n+1)}_{i} $, the Newton iteration for 
the Euler method in Eq.~(\ref{eq:Euler}) can be written in the form of a linear algebraic problem for $\mathbf \Delta_{i}^{(n+1)}$ as
\begin{align}
&\left[\mathbf{1} - h \partial_{ \mathbf V} \mathbf F(x_{i}, \mathbf V_{i}^{(n)})\right] \mathbf \Delta_{i}^{(n+1)}= \label{eq:NewtonEuler}\\ & \mathbf V_{i}^{(n)}- \mathbf V_{i-1} - h \left[ \mathbf F(x_{i}, \mathbf V^{(n)}_{i})\right], \nonumber 
\end{align}
and for the trapezoidal method in Eq.~(\ref{eq:Trapezoidal}) as
\begin{align}
&\left[\mathbf{1} - \frac{h}{2} \partial_{ \mathbf V} \mathbf F(x_{i}, \mathbf V_{i}^{(n)})\right] \mathbf \Delta_{i}^{(n+1)}= \label{eq:NewtonTrap} \\& \mathbf V_{i}^{(n)}- \mathbf V_{i-1} - \frac{h}{2} \left[\mathbf F(x_{i-1}, \mathbf V_{i-1}) + \mathbf F(x_{i}, \mathbf V^{(n)}_{i})\right], \nonumber
\end{align}
where $\partial_{ \mathbf V} \mathbf F$ denotes the  ($N+1\times N+1$) Jacobian matrix.
The improved value 
for each iteration can be obtained as 
$ \mathbf V^{(n+1)}_{i} =  \mathbf V^{(n)}_{i} - \mathbf \Delta_{i}^{(n+1)} $. At each step in the
$x$ integration, we choose the predictor as $ \mathbf V_{i}^{(0)} = \mathbf  V_{i-1}$ and iterate each 
method until the maximum relative error,  
$\text{errNewton}_i^{(n+1)} \equiv  \underset{j}{\text{Max}}\{\text{Abs}[(\mathbf  \Delta_{i}^{(n+1)})_j / (\mathbf V^{(n+1)}_{i})_j]\}$, is 
one order of magnitude smaller than the upper bound on $\text{err}_i$. We also require $\text{errNewton}_i^{(n+1)}$ to be 
smaller or equal to another accuracy control parameter, $\text{errMaxNewton}_i=0.4$, at every iteration 
step; otherwise, the stepsize $h$ is lowered in order to ensure that the Newton iteration converges.
Default values for the accuracy control parameters introduced above ($\text{err}_i$, $\text{errNewton}_i^{(n+1)}$ 
and $\text{errMaxNewton}_i$) can be adopted by changes in the settings file, see Appendix~\ref{app:settings}.

For the {\courier cBE} and {\courier fBE} routines we have implemented further, automatized 
performance optimizations. 
In the two-dimensional \textbf{cBE} case, in particular, we analytically solve in Eqs.~(\ref{eq:Euler},\ref{eq:Trapezoidal}) the 
corresponding 
number density equation for $Y_{i}$, Eq.~(\ref{Yfinalfinal}), as a function of the DM temperature 
$y_{i}$, and plug that solution into the 
DM temperature equation for $y_{i}$, Eq.~(\ref{yfinalfinal}). Effectively, this 
reduces the \textbf{cBE} system to an one-dimensional ordinary differential equation, which is how it is 
implemented in the {\courier cBE} routine.

For {\courier fBE}, the computation of the Jacobian $\partial_{\mathbf V} \mathbf F$ is implemented in vectorized form. 
Moreover, all parts in $\mathbf F$ that depend only linearly on $ \mathbf V$, e.g., the scattering operator in 
Eq.~(\ref{eq:fokkerplanckrel}), are pre-computed for every time step ($i$) and separated from the ones 
which need to be updated in every Newton step ($n$). 

Furthermore, we implemented two momentum coordinate systems, $A$ and $B$, for different 
temperature regimes, such that the phase-space density long before and long after kinetic 
decoupling remains stationary in the respective coordinates (for $Y=const.$). 
Concretely, the phase-space density is 
disretized on a uniform grid in $q_A \equiv p/\sqrt{m_{\chi}T}$ for 
$\text{Abs}[1-y^{\text{eq}}/y] <${\courier qATOqB} and $q_B \equiv (g^s_\mathrm{eff})^{-1/3}\, p/T$ 
otherwise. By default, this setting variable is set to {\courier qATOqB}$=0.1$.

For the momentum derivatives of the phase-space density in Eq.~(\ref{eq:fokkerplanckrel}) we 
implemented finite differentiation methods of higher order accuracy using several neighboring points. For 
central differentiation and fourth order accuracy implemented as a default, the boundary values of the 
phase-space density used are reflection symmetry at the origin $f_0$, introducing the ghost points to obey 
$f_{-1}(x)=f_{1}(x),f_{-2}(x)=f_{2}(x)$, and vanishing phase-space density at numerical infinity, i.e., 
$f_{N+1}(x)=f_{N+2}(x)=0$.

In practice we rely on \Mma's \lstinline$LinearSolve$ command for solving Eqs.~(\ref{eq:NewtonEuler}) 
and (\ref{eq:NewtonTrap}) in the {\courier fBE} approach. This part of the code, as well as other time-consuming matrix manipulations like collision term updates, is implemented in a C-compiled environment {(if no C compiler is 
available it is targeted to the Wolfram Virtual Machine)}, allowing for run-time performances comparable to 
pure C\texttt{++}  or Fortran code based on the analogue dgesv command from the 
LAPACK package~\cite{lapack99}.

\subsection{Validity tests}
\label{sec:test}

The evolution of the number density $Y(x)$ based on the 
traditional 
\textbf{nBE} approach were
cross-checked against results obtained with \ds, for a variety of models ranging from 
rather generic WIMP realizations to situations where DM annihilates through a narrow resonance. 
Similarly, the temperature evolution $y(x)$ computed with the kinetic decoupling (only) routines in
\ds was compared to that obtained in \db in the \textbf{cBE} and \textbf{fBE} approaches after switching 
off annihilation in our code through the settings command {\courier KDonly}$=${\courier True}. For all tested models the results were found to be in agreement at well below the percent level. 

We further checked explicitly that the number density evolution in the \textbf{cBE} and \textbf{fBE} approaches
correctly coincides with the \textbf{nBE} solution in situations where the effect of kinetic decoupling is irrelevant;
e.g.\ when the DM particles remain in full kinetic equilibrium during the freeze-out process or when the velocity dependence of the annihilation cross-section is insignificant. 

The non-trivial interference of chemical and kinetic decoupling, as taken into account in our two beyond-\textbf{nBE} approaches, was meticulously checked against independently 
implemented codes. In particular, the \db \textbf{cBE} results in the current implementation of the Scalar Singlet model (c.f.~footnote 4) were compared to results obtained from FORTRAN routines relying on the sfode~\cite{sfode} solver, 
finding agreement at the sub-percent level.\footnote{These \textbf{cBE} routines have been released as \ds 6.2.5.}
Furthermore, the \textbf{fBE} results were compared to the phase-space solver originally developed in \textsf{MatLab} based on the ODE15s solver and used in Ref.~\cite{Binder:2017rgn}.\footnote{The \textsf{MatLab} phase-space solver can be provided upon request.} 
For resonance annihilation considered in Section~\ref{sec:res} for example, {\it very} good agreement was found. Note that the \textsf{cBE} and \textsf{fBE} routines in \db do not rely on pre built-in differential equation solvers (see Section~\ref{sec:impl}), so all these tests provide an independent validity check.

A final important consistency check of the numerical \textbf{fBE} implementation concerns the question whether
the elastic scattering operator in Eq.~(\ref{eq:fokkerplanckrel}) conserves particle number.
While this is manifestly the case in the continuous limit (and hence also in \textbf{cBE}), the 
discretized version with a finite number and range in momentum can 
lead to a spurious 
change of the comoving number density if $\gamma/H \ggg 1$. Therefore some care must be taken in the 
initial high-temperature regime. One way to reduce these purely numerical artifacts is to increase the 
number of phase-space density elements, which however affects the run-time of the {\courier fBE} routine quadratically. 
We have therefore introduced an upper limit on $\gamma/H$ above which the system is assumed to be in 
kinetic equilibrium and \db adopts the \textbf{nBE} approach. By default, this upper limit is set to 
{\courier gamcap}$=10^5$.
This value and the number of phase-space elements, {\courier qN}, can be adopted in the 
settings file.\footnote{
Setting {\courier KDonly}$=${\courier True} provides a convenient way of checking for particle number
conservation in a given model. 
Note that including annihilations typically implies that particle non-conservation becomes less
of an issue at early times, since 
annihilations and creations 
further stabilizes the solution around its equilibrium value. 
}

For specific problems, optimal accuracy settings for \textbf{fBE} can differ from the default 
ones. We therefore provide for all examples in Section~\ref{sec:exampels}, as well as for the Scalar 
Singlet Model, suggested settings files. On a standard laptop computer
the time to compute the relic density for one parameter point in the
\textbf{fBE} approach can 
range from less than a second to tens of minutes (e.g.~in the presence of very narrow resonances).

Furthermore, \db contains example benchmark files with 
settings adjusted to specific cases, illustrating the usage for more challenging models. 
For further details concerning those test files see Appendix~\ref{app:timing}, 
as well as  Appendix~\ref{app:settings} for accuracy control parameters.

\section{Physics scenarios}
\label{sec:exampels}

A velocity-dependent annihilation cross-section is a necessary condition for deviations from the standard 
computation of the relic abundance, due to the interplay of chemical and kinetic decoupling. In this section 
we consider three different types of physically well-motivated scenarios where such a velocity-dependence
can appear, and where kinetic equilibrium is not necessarily maintained during chemical decoupling. 
We discuss these scenarios in some detail, both to highlight the non-trivial physics of the freeze-out 
process in at least parts of the models' parameter space, and to illustrate how \db\ can be 
used to study the freeze-out process
beyond the simplifying assumptions usually adopted in the literature.

\subsection{Resonant annihilation}
\label{sec:res}
As our first example, we consider cases where the annihilation cross-section has a strong velocity dependence induced 
by an $s$-channel resonance. The effect of early kinetic decoupling for such a setup has been studied in detail in 
Ref.~\cite{Binder:2017rgn} for the Scalar Singlet model~\cite{Silveira:1985rk,McDonald:1993ex,Burgess:2000yq}, 
where the almost on-shell particle in the $s$-channel is the SM 
Higgs. For Scalar Singlet  masses slightly smaller than half of the Higgs mass it was found that a larger coupling to the 
Higgs is required to obtain the right relic abundance, which interestingly implies that future measurements of the 
Higgs-to-invisible decay width can probe more of the parameter region than expected from the standard relic abundance 
computation, see Refs.~\cite{Duch:2017nbe,Hektor:2019ote,Abe:2020obo} for experimental probes and also for other 
Higgs resonance scenarios with similar effects.

To offer a complementary perspective, we consider here instead a generic {\it vector} mediator $A^{\mu}$ inducing 
an $s$-channel resonance {(where the parameter region close to the resonance is particularly interesting
from a model-building perspective, see, e.g., Refs.~\cite{Feng:2017drg,Bernreuther:2020koj,Fairbairn:2014aqa,Pozzo:2018anw})}. 
A minimal version of such a scenario is described by the interaction Lagrangian
\begin{align}
\mathcal{L} \supset - g_{\chi} \bar{\chi} \gamma^{\mu} \chi A_{\mu} - g_f \bar{f} \gamma^{\mu} f A_{\mu}\,, 
\label{eq:Lres}
\end{align}
allowing a Dirac fermion DM particle $\chi$ to resonantly annihilate into heat bath fermions $f$. 
The model can thus be described by a set of 5 parameters, $(m_{\chi},r,\tilde\gamma,\delta,\rho)$,
where $m_{\chi}$ is the DM mass, $r\equiv m_f/m_{\chi}$ defines the mass ratio of heat bath fermions to DM, 
$\tilde\gamma \equiv \Gamma_A/m_A$ is a dimensionless measure of the total decay width of $A^\mu$, 
$\rho\equiv\sqrt{g_{\chi} g_f}$ is the combination of coupling constants relevant for our discussion and 
$ \delta \equiv (2m_{\chi}/m_A)^2-1$ measures the deviation from the exact resonance position (at $\delta=0$).

With these definitions, the {\it annihilation} cross-section for this process, $\chi \bar{\chi} \rightarrow A^{\star} \rightarrow f \bar{f}$, can be written as 
\begin{align}
\sigma v_{\rm lab} =\frac{\rho^4 }{384\pi m_\chi^2}\frac{(1-r^2/\tilde s)^{1/2}(1+\delta)^2 }{2\tilde s -1 }  \alpha(\tilde s)D(\tilde s)\,,
\end{align}
with 
$\alpha(\tilde s)=4(2\tilde s +1)(2\tilde s +r^2)$ and $\tilde{s} \equiv s/(4m_\chi^2)$,
where $\sqrt{s}$ is the center-of-mass energy, and
the resonance is encoded in the Breit-Wigner propagator\footnote{%
For $g_\chi \gg g_f$ and $|\delta|\ll 1$, Eq.~\eqref{eq:Ds} may require a modification 
due to an energy-dependent width~\cite{Duch:2017nbe}.
}
\begin{equation}
\label{eq:Ds}
D(\tilde s) \equiv \frac{1}{\left[\tilde s (1+\delta)-1\right]^2+\tilde\gamma^2}.
\end{equation}
The {\it scattering} amplitude for $\chi f \leftrightarrow \chi f$ in the non-relativistic limit, $\tilde s\to(1+ r^2 +2 \omega/m_{\chi})/4$, 
on the other hand, is given by 
\begin{equation}
\label{eq:M2res}
|\mathcal{M}|^2_{\chi f\leftrightarrow\chi f} = \frac{16  (\delta +1)^2 \rho ^4}{\left(t(1+\delta) -4 m_\chi^2\right)^2}  \beta(\omega,t)\,,
\end{equation}
where
\begin{equation}
\beta(\omega,t) =8 m_\chi^2 \omega ^2+ 4 m_\chi^2\left(\frac{\omega}{m_\chi} +\frac12 +  \frac{r^2}{2}\right) t + t^2 \, .
\end{equation}
Following our standard convention, Eq.~(\ref{eq:M2res}) is summed over both in- and out-going spin 
states, as well as particle and anti-particle states of the scattering partners $f$. 

\begin{figure}[t]
\begin{center}
\includegraphics[scale=0.42]{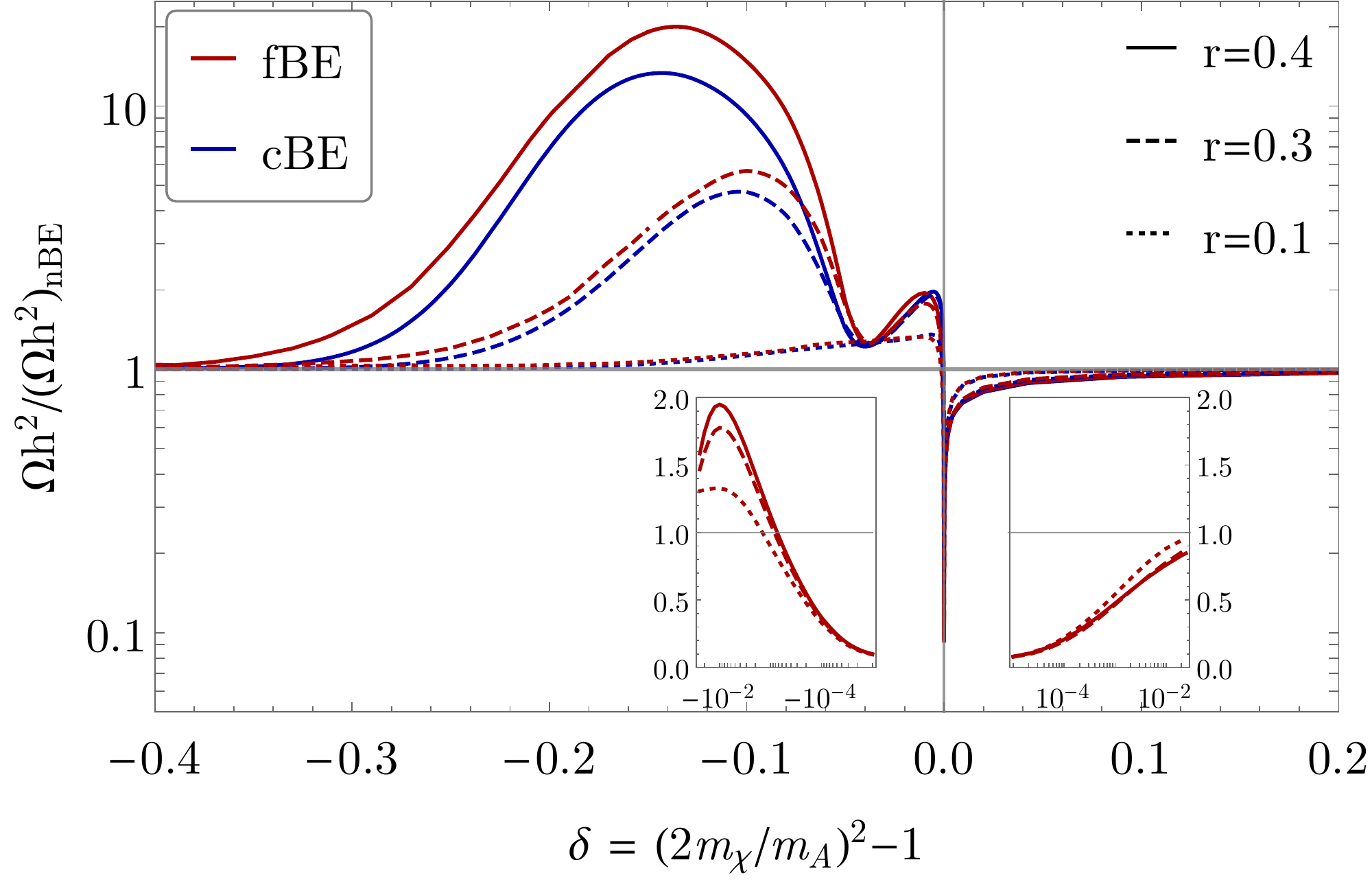}
\caption{The relic abundance, for a model with {\it resonant annihilation},  
in the \textbf{cBE} (blue) and \textbf{fBE} (red)
treatments, relative to that of the standard approach (\textbf{nBE}), as a function of the distance $\delta$ to the 
resonance. The different line style indicates different mass ratios $r=m_f/m_\chi$. 
In this example, $m_\chi=1$\,TeV, $\tilde\gamma=10^{-5}$ and coupling values are fixed by the 
requirement $2 (\Omega_{\chi} h^2)_{\rm nBE} = \Omega_{\rm DM} h^2$. The inset plots 
show zoom-ins of the resonant region around $\delta =0$. 
}
\label{fig:R1D}
\end{center}
\end{figure}

We will concentrate here on the parameter region $|\delta| < 1$ where the annihilation cross-section is 
resonantly enhanced. This implies that $\rho$ must be suppressed in order to obtain the correct relic 
abundance, leading to a corresponding suppression of the scattering amplitude. On top of that, the 
momentum transfer rate $\gamma(x)$ can be further Boltzmann suppressed for sufficiently massive 
scattering partners. In combination, these effects can lead to very early kinetic 
decoupling, interfering with the chemical decoupling process. 
In Fig.~\ref{fig:R1D} 
we quantify the impact on the relic abundance in 
such a situation, choosing for definiteness a fixed DM mass of $m_\chi=1$\,TeV and a resonance width 
of $\tilde\gamma=10^{-5}$. 
Here, we fix the coupling $\rho$ in every parameter point such that the conventional approach 
(\textbf{nBE}) matches the value of the observed DM abundance, i.e., 
$2 (\Omega_{\chi} h^2)_{\rm nBE} = \Omega_{\rm DM} h^2=0.120$.
Clearly, the \textbf{cBE} and \textbf{fBE} approaches demonstrate significant corrections to  
the standard treatment. As expected, the larger the mass ratio $r$ (from dotted to dashed to solid lines), 
the earlier DM decouples and the stronger the effect becomes.\footnote{%
We caution that our implementation of the momentum transfer rate $\gamma$ in this case, based on Eq.~(\ref{cTdef}), 
rests on the assumption of small momentum transfer during individual scattering events -- which is strictly speaking
not necessarily satisfied unless $r\ll1$.  
Still, the investigations in Appendix~\ref{app:full_cel} and the example of 
Sec.~\ref{sec:sub_threshold} show that the Fokker-Planck approximation in practice can 
work rather well even up to values of $r\sim1$.
}
Even for very light heat bath particles, however, the correction to the standard result remains sizeable 
(for $r<0.1$,  the resulting curves do not significantly differ from the $r=0.1$ case displayed
in Fig.~\ref{fig:R1D}).
The characteristic two-bump feature and the narrow dip around the exact resonance position at 
$\delta=0$ are a consequence of different DM cooling and heating effects during the chemical 
evolution that affect the DM phase-space distribution $f_\chi(x,p)$.  
The phenomenology of 
coupled chemical and kinetic evolutions here is very similar to the Scalar Singlet DM example, and 
for a detailed discussion of the origin of these features 
we refer to  to  Appendix A in Ref.~\cite{Binder:2017rgn}.

\begin{figure}[t]
\begin{center}
\includegraphics[scale=0.47]{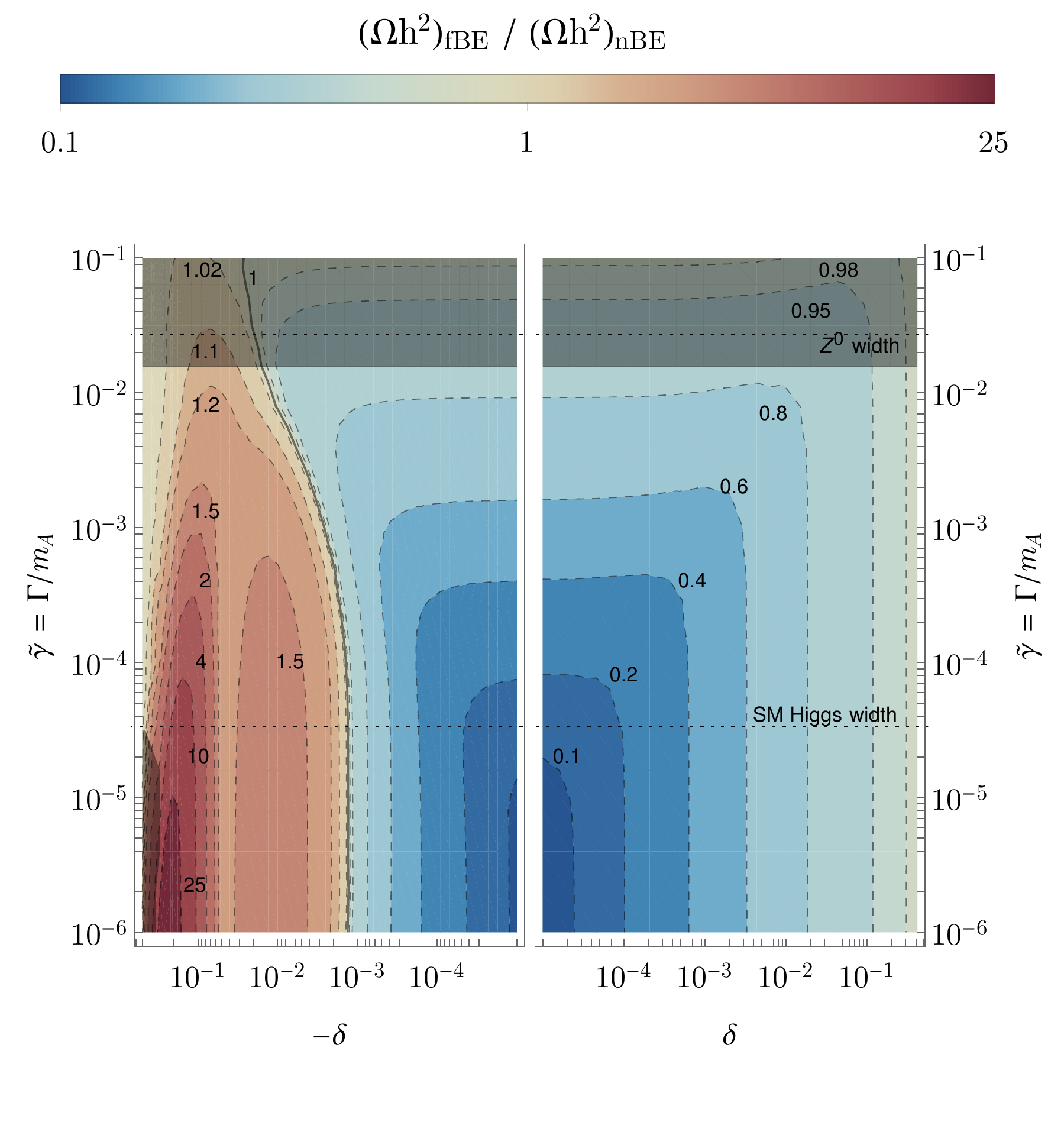}\vspace{-0.8cm}
\caption{Contour lines of constant $(\Omega_{\chi} h^2)_{\rm fBE}/(\Omega_{\chi} h^2)_{\rm nBE}$ in the $\delta-\tilde\gamma$ plane 
for a model with $ m_\chi=1$\,GeV and $r=0.3$. As in Fig.~\ref{fig:R1D}, couplings are fixed such that 
$2 (\Omega_{\chi} h^2)_{\rm nBE} = \Omega_{\rm DM} h^2$, and $\delta =(2 m_\chi/m_A)^2-1$. 
 Red (blue) colors highlight regions where the early kinetic decoupling effect overall increases (decreases) 
 the predicted relic abundance compared to the standard treatment. Dotted black lines show, for 
 comparison, the widths of the SM $Z^0$ and Higgs bosons. The gray shaded areas  
 on the edges of the 
 plot indicate parameter regions where the $\rho$ value satisfying the relic density condition cannot be 
 achieved without violating perturbativity or by extending the model (see text for more details).
 }
\label{fig:R2D}
\end{center}
\end{figure}

In Fig.~\ref{fig:R2D}, 
we complement this discussion by showing the impact on the relic abundance when 
instead varying the resonance width $\tilde\gamma$ and keeping the mass ratio $r$ fixed. 
This leads to a structure that is straightforward to relate to what is 
visible in the previous figure;
for example, the two distinctive peak regions in the bottom left corner correspond to the 
two peaks in Fig.~\ref{fig:R1D} 
(note however that here we consider a much lighter DM particle, $m_\chi=1$\,GeV, 
than in Fig.~\ref{fig:R1D}).
For most of the parameter space, a smaller  
width generally leads to a larger effect, i.e.~larger deviations from the 
standard computation. It is interesting to note that even for widths as large as that of the SM $Z$-boson 
the refined prediction of the DM abundance can deviate at a level well exceeding 
the typically quoted 
observational uncertainty of $\sim1$\% in $ \Omega_{\rm DM} h^2$ --- and hence 
at a level that would, e.g., affect 
global fits 
in a noticeable way. 

It is worth noting that for the simple model considered here, not every pair of values 
$(\delta,\tilde{\gamma})$ shown in Figs.~\ref{fig:R1D} and \ref{fig:R2D} may be a consistent choice. 
Indeed,  the minimal contribution to the width, from the interaction Lagrangian in Eq.~\eqref{eq:Lres}, is 
given by
\begin{equation}
\label{eq:gamma}
\tilde \gamma = \sum_{i=\chi,f} \frac{g_i^2}{12\pi}\left(1+ \frac{2m_i^2}{m_A^2}\right)\sqrt{1-\frac{4m_i^2}{m_A^2}}\,.
\end{equation}
In Fig.~\ref{fig:R2D} we indicate (with gray shaded regions) the values of  $(\delta,\tilde\gamma)$ that 
cannot be satisfied by Eq.~\eqref{eq:gamma} 
when $\rho=\sqrt{g_\chi g_f}$ is fixed by the relic density condition
(either because Eq.~(\ref{eq:gamma}) would imply a larger value of $\tilde \gamma$ than required, 
or because at least one of the couplings would no longer satisfy $g_\chi, g_f<\sqrt{4\pi}$,
thus indicating a breakdown of perturbativity).

\begin{figure}[t]
\begin{center}
\includegraphics[scale=0.45]{{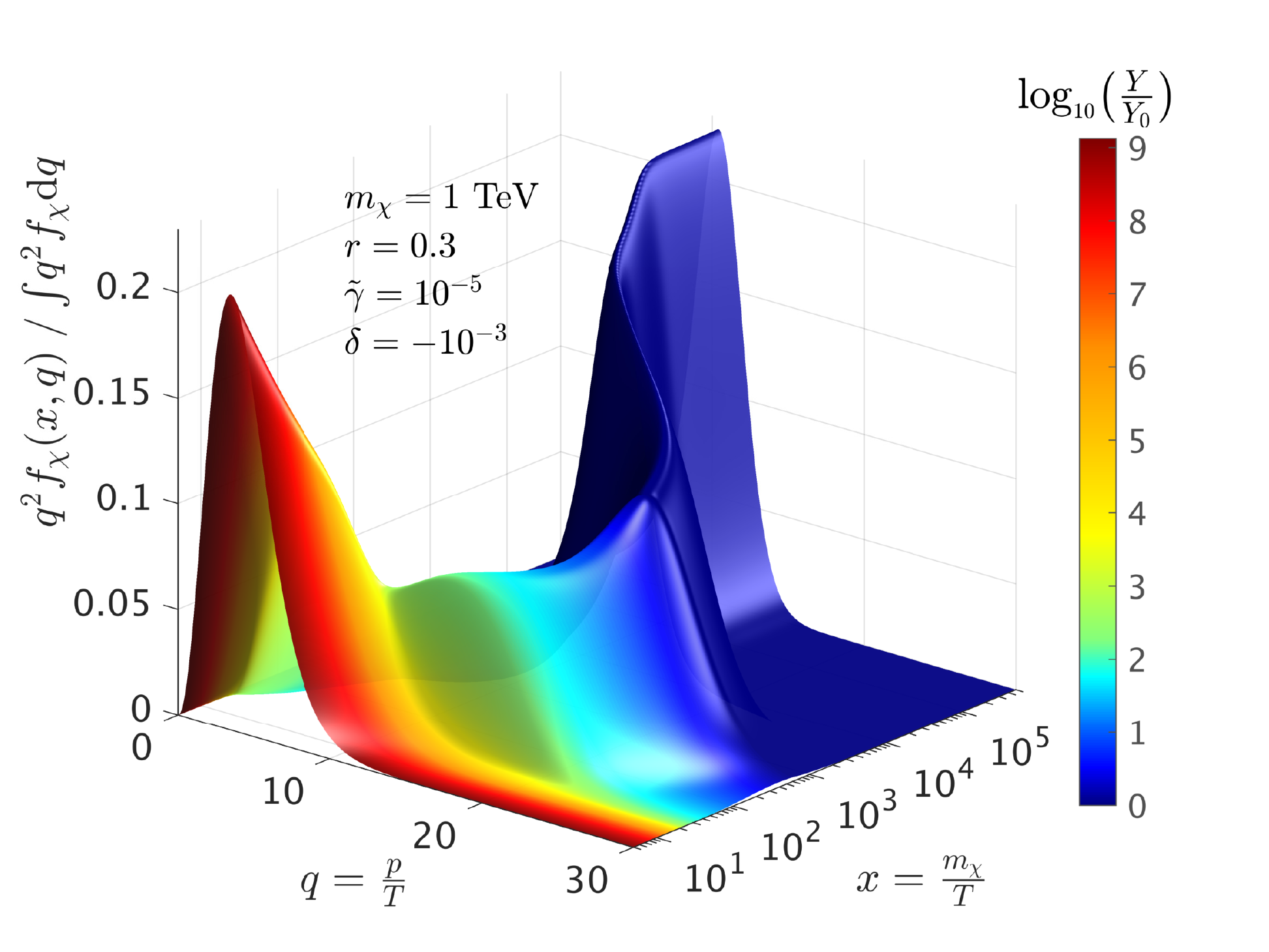}}
\caption{Time evolution of the phase-space density  $f_\chi(x,q)$, computed by the \textbf{fBE} approach. The color scale shows the DM abundance $Y(x)$, in units of its final value, which reflects the overall normalization of $f_\chi$. 
This example is for a case of DM annihilation through
a narrow resonance and an early kinetic decoupling (with model parameters stated in the legend). The coupling strength is set to {$\rho=1.13\times 10^{-2}$} in order to satisfy $2 (\Omega_{\chi} h^2)_{\rm fBE} = \Omega_{\rm DM} h^2$.
The initial equilibrium phase-space distribution is strongly distorted during chemical and kinetic decoupling, and finally remains in a highly non-thermal shape. 
}
\label{fig:R3D}
\end{center}
\end{figure}

Let us finally remark that the more narrow the resonance, the more momentum-selective is the 
annihilation process. This can lead to shapes of the distribution function that strongly differ from thermal 
ones. As a concrete example, we demonstrate in Fig.~\ref{fig:R3D} 
that the \textbf{fBE} approach implemented in \db can accurately resolve such non-trivial phase-space evolutions even for extremely 
narrow resonances (in this example a factor of a few below the Higgs width).

\subsection{Sommerfeld-enhanced annihilation}
\label{sec:sommmer}

As our second example, we consider a case where DM annihilation is Sommerfeld-enhanced
due to the presence of a light mediator. Physically, such a light mediator induces a long-range Yukawa 
potential between the DM particles that modifies their wave function, leading to a non-perturbative 
enhancement of the tree-level annihilation 
rate~\cite{Hisano:2002fk,Hisano:2003ec,Hisano:2004ds,Hisano:2006nn,Cirelli:2007xd,Mitridate:2017izz}. 
Because the Sommerfeld effect is strongly velocity-dependent, it provides a prime example of interest to 
study the interplay of chemical and kinetic 
decoupling~\cite{Dent:2009bv,Zavala:2009mi,Feng:2010zp,vandenAarssen:2012ag}. In fact, 
this is the context in which 
coupled Boltzmann equations akin to our Eqs.~(\ref{Yfinalfinal},\ref{yfinalfinal}) have first been 
proposed~\cite{vandenAarssen:2012ag}.

For simplicity, we consider the same model Lagrangian as in the previous section, Eq.~(\ref{eq:Lres}), but 
now in a very different parameter region. Concretely, we will assume $m_A \lesssim\alpha_\chi m_{\chi} $, 
 with $\alpha_\chi \equiv g_{\chi}^2/(4\pi)$, such that the vector mediator induces a long-range force. 
Furthermore, we take the heat bath particles $f$ to be (approximately) massless and only milli-charged 
under the broken $U(1)^\prime$, in the sense that $g_f \ll g_{\chi}$. 
Consequently, $\chi\bar\chi\to AA$ is the leading annihilation 
process.
The corresponding tree-level cross-section, expanded up to leading order in $v \ll 1$ 
and $m_A/m_\chi \ll 1$, is given by $(\sigma v_{\rm lab})_0 \simeq \pi \alpha^2_{\chi}/m_{\chi}^2$. 
We explicitly checked that this approximation holds for the range of parameters that we will consider here.
The full $s$-wave annihilation cross-section including the Sommerfeld effect is then obtained 
as 
\be
\sigma v_{\rm lab}=S(v_{\rm lab})\times (\sigma v_{\rm lab})_0\,,
\ee
where we use the analytical expressions for the Sommerfeld enhancement factor $S(v_{\rm lab})$ that 
were obtained, e.g., in Ref.~\cite{Cassel:2009wt, Slatyer:2009vg} after approximating the Yukawa 
potential by a Hulth\'en potential:
\be
S(v_{\rm lab}) = \frac{ \frac{\pi}{\epsilon_v} \sinh\!\Big[\frac{12\epsilon_v}{\pi \epsilon_A}\Big]}
{\cosh\!\Big[\frac{12 \epsilon_v}{\pi \epsilon_A}\Big] -\cos\!\Big[ 2 \pi \sqrt{\frac{6}{\pi^2 \epsilon_A}\!-\!\left( \frac{6 \epsilon_v}{\pi^2 \epsilon_A}\right)^2}\Big]},
\ee
where $\epsilon_v \equiv v_\text{lab} /(2 \alpha_\chi)$ and $\epsilon_A \equiv m_A/(\alpha_\chi m_\chi)$. 

We concentrate on the parameter region where $g_f$ is (just) large enough for the mediator to be in equilibrium during freeze-out, established by the decay and inverse decay processes $A \leftrightarrow f \bar{f}$. 
Requiring the decay rate to satisfy $\Gamma_{A \rightarrow f \bar{f}}/H \gtrsim 1 $ for $T \lesssim m_{\chi}$, one finds that this is realized for $g_f \gtrsim 3 \times 10^{-6} (0.1/r_A) (m_{\chi}/\text{TeV})^{1/2}$,%
\footnote{Since $m_A \ll m_{\chi}\sim T$ we need to take into account relativistic corrections to the decay rate in the form of a time dilation factor $m_A/E$, see, e.g., Ref.~\cite{KAWASAKI1993671}. We estimate the decay rate for $ m_A \ll T$  as $\Gamma_{A \rightarrow f \bar{f}} \simeq \Gamma_{A \rightarrow f \bar{f}}^0 m_A/(2T)$,  where $\Gamma_{A \rightarrow f \bar{f}}^0=g_f^2 m_A/(12 \pi)$ is the decay rate in the $A$ rest frame and $m_A/(2T)$ is the average time dilation in the plasma frame. In this estimate, Pauli blocking of final states was neglected and we adopted a classical Maxwell-Boltzmann statistics for the vector boson~$A$.} %
where $r_A\equiv m_A/m_{\chi}$. 
For simplicity we will exclusively focus on coupling values that are not significantly larger than this lower bound,
thus implementing the earliest possibility when DM can kinetically decouple in this model.
In this case, Coulomb-like scattering processes 
$f \chi \leftrightarrow f \chi$ can be neglected -- as we checked explicitly -- and DM is kept in kinetic 
equilibrium by the Compton-like scattering with the vector particle, $\chi A \leftrightarrow \chi A$.
The corresponding scattering amplitude is for non-relativistic DM given by
%
%
\be
\label{eq:SommerScatter}
|\mathcal{M}|^2_{\chi A\leftrightarrow\chi A} =   \frac{8 g_{\chi}^4 \beta(\omega,t)}{(2 m_{\chi} \omega + m_A^2)^2 (2 m_{\chi} \omega - m_A^2 + t)^2}\,,
\ee
where, to leading order in the DM mass,
\begin{align}
\beta(\omega,t)&\simeq 4 m_{\chi}^4 (4 m_A^4 - 4 m_A^2 t + 8 \omega^4 + 4 \omega^2 t + t^2).
\end{align}
In this parameter region, the phenomenology of the model thus only depends on the coupling $g_\chi$ 
(unlike in the previous section, where the relevant coupling combination was $\rho=\sqrt{g_\chi g_f}$).
For relevant coupling values and the scattering amplitude as in Eq.~(\ref{eq:SommerScatter}), 
kinetic decoupling occurs once the vector boson $A$ enters the non-relativistic regime, 
causing a Boltzmann suppression of the momentum transfer rate $\gamma(x)$.

\begin{figure}[t]
\begin{center}
\includegraphics[scale=0.42]{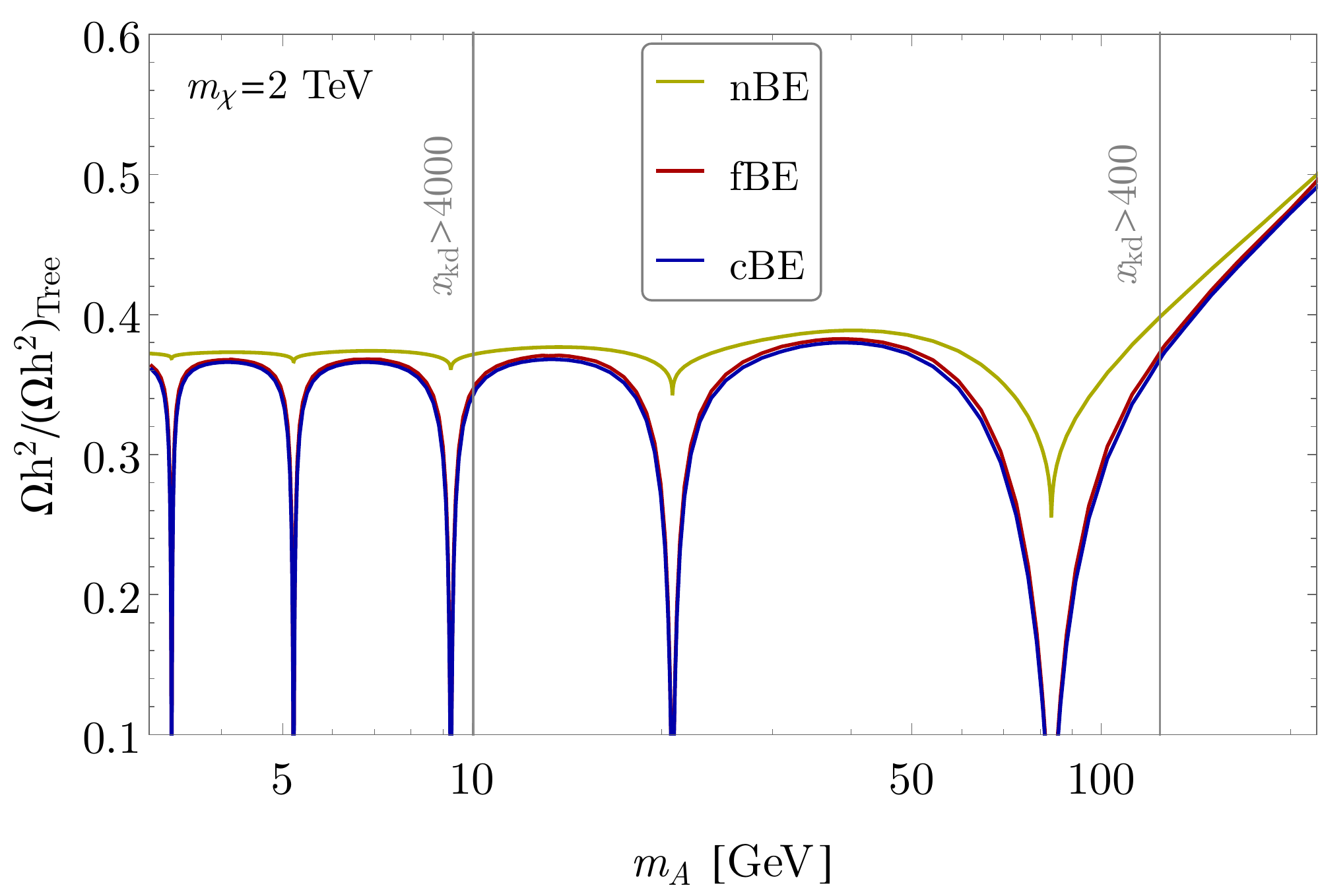}
\caption{Relic abundance for the \emph{Sommerfeld-enhanced annihilation} example, for a fixed DM mass
of $m_{\chi}=2$\,TeV and as a function of the mediator mass $m_A$. The coupling is fixed to 
$\alpha_\chi = 0.07$ 
from the requirement $2 (\Omega_{\chi} h^2)_{\text{Tree}} = \Omega_{\text{DM}}h^2$, based on the 
(velocity-independent) tree-level annihilation cross-section $(\sigma v_{\rm lab})_0$ without Sommerfeld 
enhancement. The predictions from the traditional \textbf{nBE} computation clearly demonstrate the 
impact of the Sommerfeld effect on the annihilation rate, while the improved  calculations based on 
\textbf{cBE} and \textbf{fBE} 
quantify the additional impact of kinetic decoupling on the relic abundance.}
\label{fig:SE1}
\end{center}
\end{figure}

\begin{figure}[t]
\begin{center}
\includegraphics[scale=0.37]{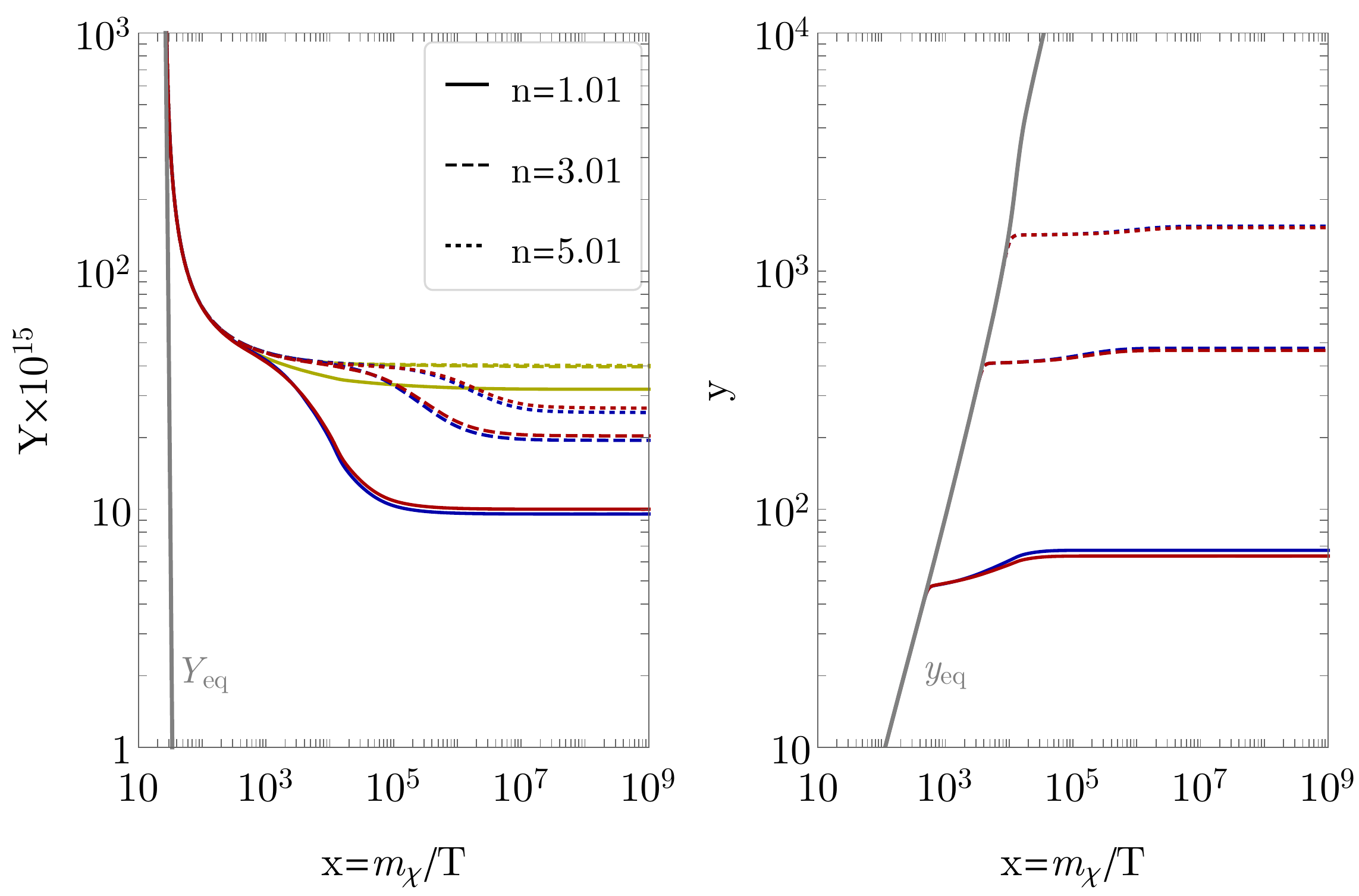}
\caption{Evolution of DM abundance $Y(x)$ (left) and velocity dispersion $y(x)$ (right) for 
the Sommerfeld model, with $\alpha_\chi $ and $m_\chi$ as in Fig.~\ref{fig:SE1}. Line colors correspond 
to \textbf{nBE} (green), \textbf{cBE} (blue) and \textbf{fBE} (red). Line styles refer, as indicated, to different
values of $m_A$, chosen such that $n^2= 6 \alpha_{\chi} m_{\chi}/(\pi^2  m_A)$ is close to integer 
(parametric resonance condition). 
The second annihilation era characteristic for these models is clearly visible as a drop of $Y$ in the left 
panel; the corresponding increase in the DM temperature visible in the right panel is caused by 
a self-heating due to efficient annihilation out of thermal equilibrium.}
\label{fig:SEYy}
\end{center}
\end{figure}

In Fig.~\ref{fig:SE1}
we show the resulting relic density in this setup, for a range of mediator masses $m_A$ and
fixed DM mass ($m_\chi=2$\,TeV) and coupling ($\alpha_\chi=0.07$). As is clearly visible in this figure, the 
Sommerfeld effect alone affects the relic density by a factor of 2--3 compared to the tree-level 
expectation. For selected values of mediator masses, the difference between the standard approach and 
both our beyond-\textbf{nBE} approaches can be even larger. For these parameter combinations, bound 
states with zero binding energy can form; parametrically, these exist precisely at threshold for 
$m_A/(\alpha_{\chi} m_{\chi})= 6/(n^2 \pi^2)$ in the case of the Hulth\'en potential (with $n \in \mathbb{Z}^+$). 
Close to these parametric resonances the Sommerfeld effect scales as $S(v_{\rm lab})\propto1/v_{\rm lab}^2$ for $v_{\rm lab} \lesssim m_A/m_{\chi}$,
leading to a second period of annihilation after kinetic 
decoupling~\cite{Dent:2009bv,Zavala:2009mi,Feng:2010zp,vandenAarssen:2012ag}.
We illustrate this in Fig.~\ref{fig:SEYy} 
by plotting the evolution of the DM abundance $Y(x)$ for three selected
mediator masses close to such a parametric resonance (left panel). Due to the 
inverse velocity dependence of $S(v_{\rm lab})$, furthermore, DM particles prefer to annihilate at smaller momenta, 
leading to a self-heating after kinetic decoupling and hence an increase in $y(x)$ (right panel)
-- see also the discussion in Ref.~\cite{vandenAarssen:2012ag}.

Inspecting Fig.~\ref{fig:SE1}, we find significant deviations from the standard computation also for 
parameter regions further away from the exact resonance condition. 
Here, the Sommerfeld effect scales only as $S(v_{\rm lab})\propto1/v_{\rm lab}$, which correspondingly leads to a significantly weaker re-annihilation effect. 
Let us point out that even
for situations with a relatively early
kinetic decoupling as studied in this section (though still much later than what we studied in Section 
\ref{sec:res}), the difference between \textbf{cBE} and \textbf{fBE} is 
as expected rather small, independently of the vicinity to a parametric resonance. 
This holds both for the evolution of  the number density and that of the DM temperature,
as also demonstrated earlier in Ref.~\cite{Binder:2017lkj}.

In general, the effect on 
the DM abundance due to 
kinetic decoupling is of course larger for earlier 
kinetic decoupling (assuming all other parameters to be fixed).
However, we remark that even for later kinetic decoupling than in the minimal case discussed here -- e.g. 
in the case where Coulomb-like scattering dominates, for much larger coupling values $g_f$ -- the correct 
prediction of $\Omega_\chi h^2$ can be sensitive to the interplay of kinetic and chemical freeze-out 
processes, making it necessary to go beyond the traditional \textbf{nBE} approach.

\subsection{Sub-threshold annihilation}
\label{sec:sub_threshold}

As our final example we consider a situation where the annihilation process that sets the relic density 
is kinematically not accessible in the limit of vanishing velocities, sometimes dubbed `forbidden' 
DM~\cite{DAgnolo:2015ujb}.

For concreteness, we adopt a simple scenario with an interaction Lagrangian given by 
\be
\label{eq:L_threshold}
\mathcal{L} \supset -\frac{\lambda}{4} \phi_1^2 \phi_2^2 +y_f \phi_2\bar f f\,,
\ee
where the scalar $\phi_1$ takes the role of the DM particle. It directly interacts only with 
the scalar $\phi_2$, which we assume to be close in mass with $\phi_1$, i.e., $r\equiv m_2/m_1\sim 1$,
and to be in thermal contact with the heat bath fermions $f$. 
To lowest order, the total DM annihilation cross-section is determined by the process $\phi_1\phi_1\to\phi_2\phi_2$, 
and given by
\begin{align}
(\sigma v_{\rm lab})^{2\to2} =  \frac{\lambda^2}{32 \pi }  \frac{\sqrt{1-4m_2^2/s}}{s-2 m_1^2}\,. 
\label{eq:sv_threshold}
\end{align}
For $r>1$, this process is only open to particles in the high-energy tail of the DM distribution,
leading to an exponential suppression of $\langle\sigma v\rangle$ and a corresponding increase of $\Omega_{\chi} h^2$.

Similar to the Sommerfeld enhancement example discussed in the previous section, we restrict our 
discussion for simplicity to couplings $y_f$ (just) large enough to bring $\phi_2$ into equilibrium before the 
onset of the freeze-out 
process, through (inverse) decays $\phi_2 \leftrightarrow \bar f f$. Demanding for concreteness
$\Gamma_{\phi_2 \leftrightarrow \bar f f}/H \gtrsim 1$ for $T \lesssim m_{1}$, this requires 
$y_f \gtrsim 10^{-7}\,r^{-\frac12} \left(m_1/{\rm TeV}\right)^{\frac12}$. 
At the same time we require for consistency that $y_f$ is small enough such 
that {\it i)} close to the threshold at $r\sim 1$, 3-body processes of the form 
$\phi_1\phi_1\to\phi_2\phi_2^*\to\phi_2 f\bar f$ can be 
neglected and that {\it ii)} the loop-induced scattering with $f$ is subdominant compared to the tree-level 
scattering  with the Boltzmann-suppressed $\phi_2$ particles, via
$\phi_1 \phi_2 \leftrightarrow \phi_1 \phi_2$.\footnote{%
We explicitly checked that these requirements actually allow values of $y_f$ up to a few orders of 
magnitude above the lower bound due to the thermalization condition, with condition {\it i)} being the more 
stringent requirement. Concretely, we followed Ref.~\cite{Bringmann:2017sko} to compute (for $m_f\ll m_2$)
\be
\sigma v_{\rm lab}^{2\to3}\simeq\frac{g^2/(8\pi)}{ s-2 m_1^2}
\int_0^{\tilde\mu}\!\!\frac{d\mu}{\pi}
\frac{\tilde\gamma\mu}{(\mu\!-\!1)^2\!+\!\tilde\gamma^2}\,\lambda^\frac12\!\!\left(\!1,\frac{m_2^2}{s},\frac{\mu m_2^2}{s}\!\right).
\ee
Here, $\lambda$ is the K\"all\'en (or triangle) function, $\tilde\mu\equiv(1-\sqrt{s}/m_2)^2$ and 
$\tilde\gamma\equiv \Gamma_{\phi_2\to\bar ff}/m_2= y_f^2/(8\pi)$. The total annihilation cross-section
entering the Boltzmann equation, taking into account double-counting issues, is thus obtained as
$ \sigma v_{\rm lab}=\sigma v_{\rm lab}^{2\to3}-\sigma v_{\rm lab}^{2\to2}$ 
(which correctly reduces to $\sigma v_{\rm lab}\simeq\sigma v_{\rm lab}^{2\to2}$ for $s\gtrsim4m_2^2$).
} 
In this parameter region, the momentum transfer rate, Eq.~(\ref{eq:fokkerplanckrel}), is then given by
\begin{align}
\gamma(x) \approx m_{1} \frac{\lambda^2}{6 \pi^3} \frac{r^2}{(1+r)^4}  x^{-2} e^{-rx} \,.
\label{eq:thFP}
\end{align}
We note that this expression explicitly features the already mentioned exponential suppression due to 
scattering with non-relativistic $\phi_2$ particles. Hence, we expect kinetic decoupling to happen very early
in this model, around the time of chemical decoupling.

As already stressed in Section \ref{sec_collision}, however, the approximate scattering collision term in 
Eq.~(\ref{eq:fokkerplanckrel}), including the momentum transfer rate, relies on the assumption of small 
momentum transfer compared to the average DM momentum -- which is typically {\it not} satisfied for 
scattering partners that are close in mass to the DM particle.
For the simple case of a constant scattering amplitude, consistent with our interaction Lagrangian in 
Eq.~(\ref{eq:L_threshold}), we demonstrate in Appendix \ref{app:full_cel} that it is possible to instead
implement the scattering collision term $C_{\text{el}}$ without this assumption. Concretely, we perform analytically all 
integrals for $W$ in Eq.~\eqref{eq:Wmain}
, c.f.~Eq.~(\ref{eq:WTH}), and use this expression to compute 
$C_\text{el}$ in Eq.~(\ref{eq:Cel}). This allows us to contrast our results to those based on the 
approximate scattering term $C_\text{FP}$ (relying on small momentum transfer).

\begin{figure}[t]
\centering
\includegraphics[scale=0.42]{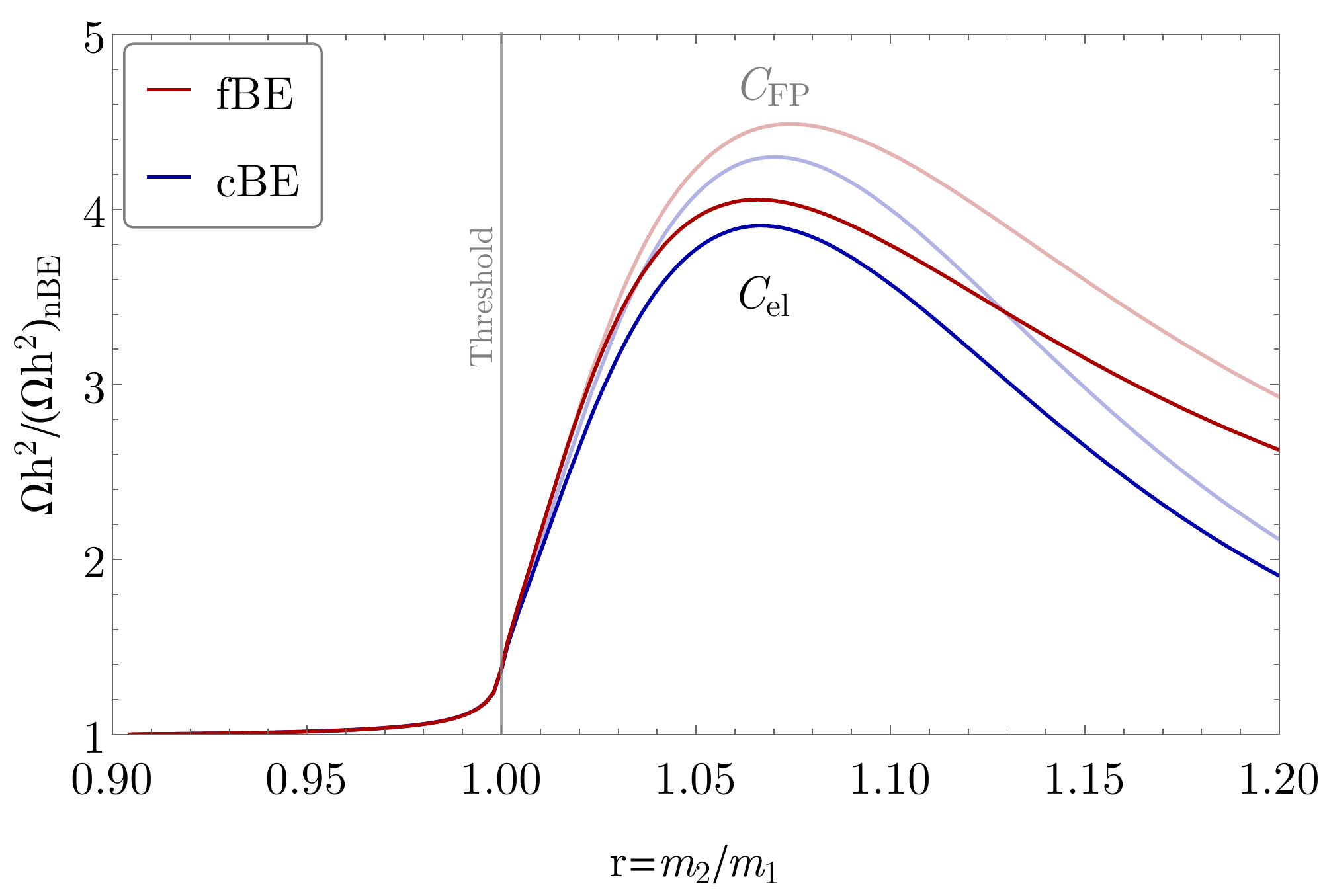}
\caption{Relic abundance for the \emph{sub-threshold annihilation} example, for a fixed DM mass
of $m_{1}=100$\,GeV and in function of the final state mass ratio $r=m_2/m_1$. 
The value of the $\phi_1-\phi_2$ coupling, $\lambda$, is determined by the requirement 
$(\Omega_{\chi} h^2)_{\text{nBE}} = \Omega_{\text{DM}}h^2$;
in the `forbidden' region, for $r>1$, $\lambda$ is thus exponentially sensitive to $r$.
Solid red (blue) lines correspond to results from the \textbf{fBE} (\textbf{cBE}) treatment based on the full 
scattering term of this example, while the corresponding results based on the small-momentum transfer 
approximation in Eq.~(\ref{eq:fokkerplanckrel}) are plotted with lighter shading.
}
\label{fig:thscan}
\end{figure}

In Fig.~\ref{fig:thscan} 
we plot the relic density that results from the \textbf{fBE} and \textbf{cBE} approaches, 
with different implementations of the scattering terms, relative to that from the \textbf{nBE} 
approach. Here, for each value of $r$, the coupling $\lambda$ is fixed by the requirement that the standard \textbf{nBE} 
prediction matches the observed abundance. 
As is clearly visible, early kinetic decoupling can have a significant impact 
on the relic abundance also for this threshold example, at least for $r \gtrsim 1$.
While it is remarkable that both scattering term prescriptions show
qualitatively the same result, however, in this model DM  turns out to be kept slightly more efficiently in kinetic equilibrium with the 
more correct $C_{\text{el}}$ approach; see also Fig.~\ref{fig:NR2mom} in Appendix~\ref{app:full_cel}.
This explains why the $C_{\text{el}}$  prescription leads to a relic density slightly closer to that
of the standard \textbf{nBE} approach than what we find with the Fokker-Planck approximation.

\begin{figure}[t]
\centering
\includegraphics[scale=0.37]{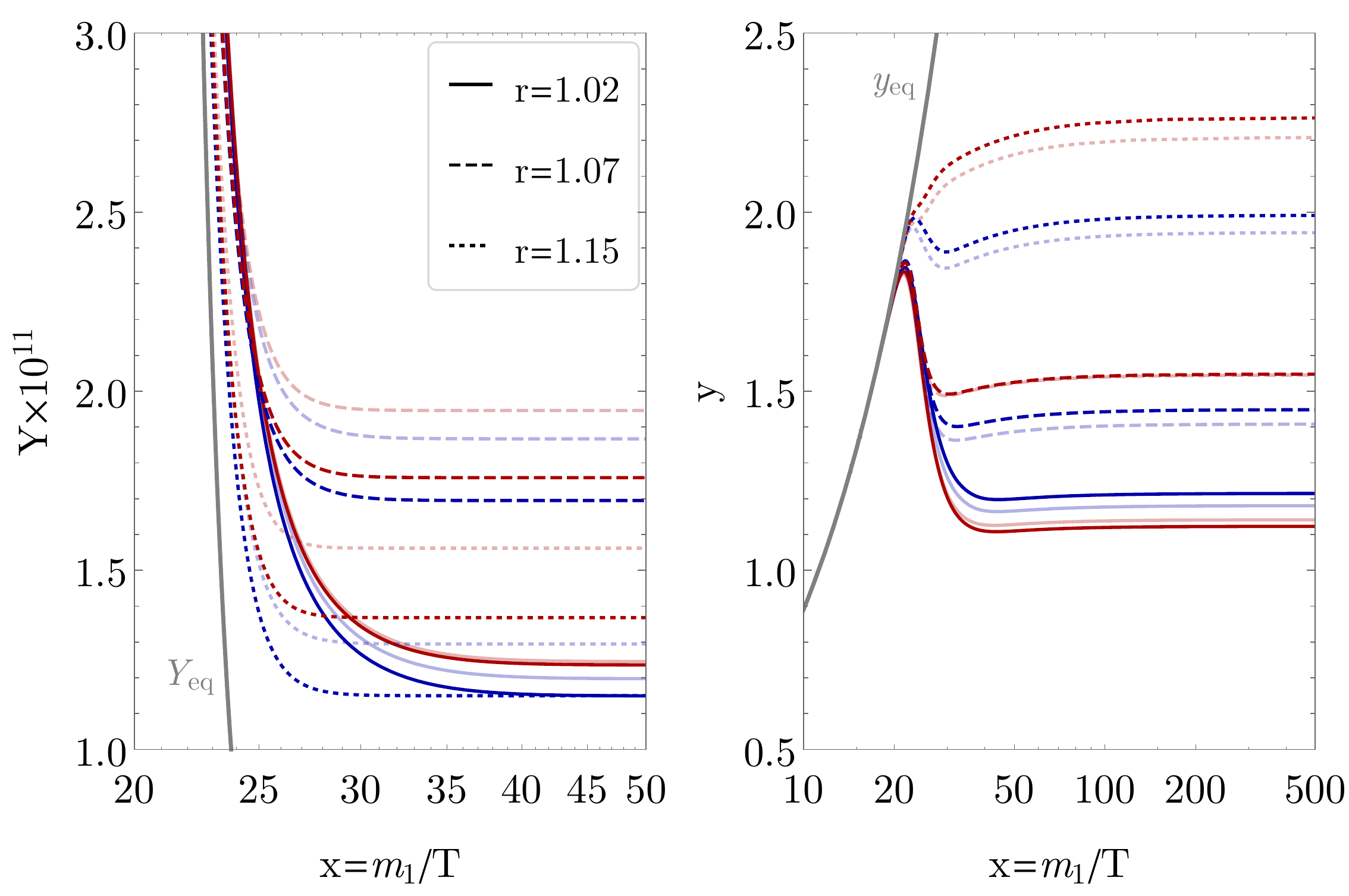}
\caption{Evolution of DM abundance $Y(x)$ (left) and velocity dispersion $y(x)$ (right)
for the sub-threshold model with three selected values of $r$. 
DM mass and couplings are fixed as in Fig.~\ref{fig:thscan}, and line styles chosen in accordance with 
that figure. For these threshold examples, DM needs high momenta for annihilation to be kinematically 
allowed, which results in a phase of cooling around and after kinetic decoupling, and hence a drop in $y(x)$.
}
\label{fig:thYy}
\end{figure}

In Fig.~\ref{fig:thYy}, 
we show the abundance and temperature evolution for selected values of $r>1$.
Qualitatively, DM needs higher momenta to overcome the annihilation threshold, leading to a self-cooling 
phase as soon as it is no longer (fully) kinetically coupled to the particles $\phi_2$ that remain in 
equilibrium with the heat bath.
Due to this cooling, annihilation becomes less efficient earlier, resulting in a higher DM abundance than in 
the standard computation.

Let us close this section by briefly mentioning again 
that the kinematically suppressed annihilation cross-section implies that the coupling $\lambda$ must grow exponentially with $r$ in order to maintain the 
correct DM abundance. For example, in the \textbf{nBE} approach with $m_1=100$\,GeV very large 
couplings $\lambda \gtrsim 4\pi$ are  reached at $r\gtrsim 1.2$; for the \textbf{cBE} and \textbf{fBE} 
approaches this happens at even smaller values of $r$, closer to the peak visible in Fig.~\ref{fig:thscan}.

\section{Summary}
\label{sec:summary}

Thermal freeze-out is widely considered one of the most intriguing mechanisms for the production of DM.
The underlying assumption of by far most relic density calculations in the literature 
is that  kinetic equilibrium is maintained during the freeze-out. Here we have presented a new public tool, \db,
to explicitly gauge the impact of this assumption in cases where it might not be valid.  
To do so, the code offers various alternative schemes to calculate the relic density
that take into account the intriguing effects of kinetic decoupling during the chemical freeze-out process, including a full 
calculation at the phase-space level.

In fact, it has repeatedly been pointed out that chemical and kinetic decoupling can be intertwined in a 
way that significantly affects the result of relic density calculations~\cite{vandenAarssen:2012ag,Duch:2017nbe,Kamada:2017gfc,Berlin:2018tvf,Kamada:2018hte,Abe:2020obo,Binder:2017rgn,Brummer:2019inq,Ala-Mattinen:2019mpa,Kuflik:2015isi,Kuflik:2017iqs,Fitzpatrick:2020vba,Biswas:2020ubd,Croon:2020ntf}. To provide context, we 
have therefore devoted a large part of this article (Section \ref{sec:exampels}) to a comprehensive and
updated study of three physically well-motivated scenarios of annihilating DM where this is 
the case, illustrating the need 
to move beyond the standard treatment and demonstrating that this is possible without compromising 
on accuracy. 
This is clearly  important for global fits that include parameter regions in their scans where the 
interplay between kinetic and chemical equilibrium cannot be neglected, but can turn out to be 
relevant also in various model-building considerations.

Let us finally stress that, while the main focus of \db is the relic density, its output is by no
means restricted to a single number. Rather, it allows to compute the full time evolution of the
DM phase-space density, or its lowest moments, which may be connected to further late-time
observables.  For example, a non-standard velocity distribution would
affect how free streaming impacts the matter power spectrum of density 
perturbations~\cite{Boyanovsky:2010pw,Hager:2020une}, which is one of the main
motivations for why 
linear perturbation solvers like 
\href{https://lesgourg.github.io/class_public/class.html}{\textsf{CLASS}}~\cite{Lesgourgues:2011rh} explicitly support 
externally tabulated (non-standard) phase-space densities.
An exploration of such effects in the context of kinetic decoupling interfering with chemical decoupling 
would be warranted, but beyond the scope of this work.

\vfill
\begin{acknowledgements}
We would like to thank Tomohiro Abe and Ayuki Kamada for discussions about our results of the exact scattering collision term treatment vs. Fokker-Planck approximation. T.~Binder was supported by World Premier International Research Center Initiative (WPI), MEXT, Japan,
by the JSPS Core-to-Core Program Grant Number JPJSCCA20200002 and by JSPS KAKENHI Grant Number 20H01895.
He would also like to thank J. Hisano and S. Matsumoto for the additional support.
M.G.\ acknowledges partial support from the European Unions Horizon 2020 research and innovation program under grant agreement No 690575 and No 674896.
A.H. is supported in part by the National Science Centre, Poland, research grant No. 2018/31/D/ST2/00813.
\end{acknowledgements}

\newpage
\appendix

\renewcommand{\appendixname}{\!\!}
\section*{Appendices}
\addcontentsline{toc}{section}{Appendices}

\section{Getting started}
\label{app:quick_start}

\db can be downloaded as a zipball from \href{https://drake.hepforge.org/}{drake.hepforge.org}. The only prerequisite for using it is to have \Mma or the free \WE installed. 

\subsection{Quick start}
\label{app:quick}

Unpack the zipball and open the \Mma notebook \textbf{main.nb} located in the main directory. This example program demonstrates the loading of the \db package, code usage, printing routine output in form of plots, and saving the results. In the following, we expand on the description of all the consecutive steps in this example program.

To load the package, start a fresh kernel and
execute
\lstset{upquote=true}
\begin{lstlisting}
SetDirectory["<path>"];
Needs["DRAKE`"]
\end{lstlisting}
\lstset{upquote=false}
where \lstinline$<path>$ is the path to the main directory. From this point on, all new symbols 
can be listed with 
\lstset{upquote=true}
\begin{lstlisting}
Names["DRAKE`*"]
\end{lstlisting}
\lstset{upquote=false}
and documentation of the main functions and variables are provided with the usual syntax, e.g., 
calling \lstinline$?cBE$ returns information about the {\courier cBE} routine. 
 
All input quantities can be loaded with the command
\begin{lstlisting}
GetModel["<model>", "<parameters>", "<settings>"];
\end{lstlisting}
The three arguments are names of files (without their extension .wl), located in the sub-directory \lstinline$./models/<model>/$. 
\db  comes with a set of pre-implemented models, where \mbox{\lstinline$<model>$} $\in \{$\lstinline$ScalarSingletDM$, \lstinline$WIMP$,  \lstinline$VRES$, \lstinline$SE$, \lstinline$TH$$\}$. The first element in this 
set is the Scalar Singlet DM model, the second is a 
WIMP-like test model, while the last three are the scenarios presented in 
Section~\ref{sec:exampels}. For any of these models, choose one of the parameter files 
\lstinline$<parameters>$ $\in \{$\lstinline$bm1$, \lstinline$bm2$,  \lstinline$bm3$$\}$, 
implementing different benchmark values of model parameters.
Correspondingly, the files \lstinline$<settings>$$\in \{$\lstinline$settings_bm1$, \lstinline$settings_bm2$,  \lstinline$settings_bm3$$\}$ contain suggested values of
accuracy control parameters and Boolean variables for run options.

After these initialization steps, the main \db routines are ready to perform $\Omega_{\chi} h^2$ 
computations. For results obtained with the  \textbf{nBE} approach, e.g., this amounts to calling
\begin{lstlisting}
PrepANN;
nBE
\end{lstlisting}
for any of the pre-implemented models. 
In the default usage with the Fokker-Planck approximation for $C_{\text{el}}$, 
on the other hand,
the \textbf{cBE} approach consists of the calls 
\begin{lstlisting}
PrepANN;
PrepANN2;
PrepSCATT;
cBE
\end{lstlisting}
and the default \textbf{fBE} approach consists of the calls 
\begin{lstlisting}
PrepANNtheta;
PrepSCATT;
fBE
\end{lstlisting}
When working with all approaches simultaneously, it is sufficient to call each \emph{preparatory} 
routine  once.
Optional procedures, where e.g.\ \lstinline$PrepANNtheta$ and \lstinline$PrepSCATT$ are omitted, are explained in Section~\ref{app:settings}.

The output of the  \emph{preparatory}  routines and Boltzmann \emph{solvers} is summarized in Table~\ref{tab:primaryclassroutines}. 
\begin{table}
\begin{center}
\begin{tabular}{l||l|l} 
\toprule 
   Call & Output & Output names \\ \midrule
    \lstinline$PrepANN$ & $\langle \sigma v\rangle$ & \courier{tsv},\courier{svx[x]}   
     \\ \midrule 
    \lstinline$PrepANN2$  & $\langle \sigma v\rangle_2$ &\courier{tsv2},\courier{sv2x[x]} 
       \\ \midrule 
    \lstinline$PrepANNtheta$  & $ \langle \sigma v\rangle_{\theta}$ &\courier{tsvtheta},   \\ 
&  & \courier{isvtheta} \\ \midrule
\lstinline$PrepSCATT$  & $ \gamma(x) $& \courier{tgamma},\courier{gam[x]}   \\ \midrule \midrule
   
\lstinline$nBE$  & $(\Omega_{\chi} h^2)_{\text{nBE}}$ &  \courier{Oh2nBE}  \\ 
    
    &  $Y_{\rm nBE}(x)$ & \courier{tYnBE}=\{x,$Y_{\rm nBE}$\}  \\ \midrule 
    
    \lstinline$cBE$ & $(\Omega_{\chi} h^2)_{\text{cBE}}$ & \courier{Oh2cBE} \\ 
    &  $Y_{\rm cBE}(x)$ & \courier{tYcBE}=\{x,$Y_{\rm cBE}$\} \\ 
    &  $y_{\rm cBE}(x)$ & \courier{tycBE}=\{x,$y_{\rm cBE}$\} \\ \midrule 
    \lstinline$fBE$  & $(\Omega_{\chi} h^2)_{\text{fBE}}$ & \courier{Oh2fBE} \\ 
&  $  Y_{\rm fBE}(x)$ &  \courier{tYfBE}=\{x,$Y_{\rm fBE}$\}\\
&  $ y_{\rm fBE}(x) $ & \courier{tyfBE}=\{x,$y_{\rm fBE}$\} \\
&  $ q^2 f_{\chi}(x,q)$ & \courier{tfDM}=\{x,\{$q^2 f_\chi$\}\}\\ \bottomrule
\end{tabular}
\end{center}
\caption{Output summary of \db routines. 
The last column lists the main output variables, tables, and functions stored in the kernel session memory after calling the routine. Cross-section outputs are in units of $\text{GeV}^{-2}$ and $\gamma$ in GeV. All output names starting with `t' are tables and names with an [x] dependence are interpolating functions based on these tables. The tables of the Boltzmann solvers contain all computed $Y$ and $y$ values from initial to final $x$ except {\courier{tfDM}}, which contains for memory reason only snapshots of the unity-normalized phase-space density at the initial, intermediate (when switching from momentum coordinates $q_A$ to $q_B$, see Section~\ref{sec:impl}) and final $x$ value. The preparatory routines are stored in the file \textbf{rates.wl} and the Boltzmann solvers in \textbf{nBE.wl}, \textbf{cBE.wl}, and \textbf{fBE.wl} files in the \db source directory.
}
\label{tab:primaryclassroutines}
\end{table}
All listed output variables are stored in the kernel session memory and can be directly accessed and saved to a file \lstinline$<outfile>$ with, e.g., \db's command \lstinline$save["<outfile>"]$.

Values of any model or setting parameters can be changed actively 
during a notebook session. This makes scans over, e.g., the DM mass simple to perform. 
Alternatively, one can create different \lstinline$<parameters>$ and \lstinline$<settings>$ files 
and repeatedly use \lstinline$GetModel$.

\subsection{Template script}
\label{app:tempscript}

The release also contains a template script \textbf{main.wls}, which serves as a streamlined and 
customizable way to use \db directly from a terminal. For any of the five pre-implemented models 
(see Section~\ref{app:quick} for file names), in particular, the template script can be executed as
\begin{lstlisting}
> wolframscript main.wls <model> <parameters>     <settings>
\end{lstlisting}
This loads \db, executes \lstinline$GetModel$ and runs the \textbf{nBE}, \textbf{cBE}, and \textbf{fBE} approaches with a minimum amount of routine  calls and according to the user's option settings (discussed in Section~\ref{app:settings}).
Figure~\ref{fig:code}, in Section~\ref{sec:code}, graphically illustrates its workflow. Results stored in the session's memory are then saved to a file, with name and path specified in the \lstinline$<settings>$ file. Important results are  also directly displayed in text format, along with the computation time. The template script can also be called in a notebook environment with a convenient wrapper \lstinline$RunMain$, as demonstrated in \textbf{main.nb}.

\subsection{Tests}
\label{app:timing}

The release contains a test notebook, \textbf{test.nb}, located in the \lstinline$"./test"$ sub-directory.
It can be used for checking the \db installation against pre-computed results.
While for the simplest pre-implemented model (WIMP) the calculation takes seconds, it can take up to several minutes 
for the most complicated benchmark scenario (VRES, close to a narrow resonance). The computational 
time of the Boltzmann solvers is typically less than it takes for the corresponding preparatory routine calls. 

For \WE users, move to the  \lstinline$"./test"$ directory and run the test script by, e.g.,
\begin{lstlisting}
> wolframscript test.wls WIMP bm_WIMP settings_WIMP
\end{lstlisting}
which displays the test results for the WIMP model.

\subsection{Adding a new model}
\label{app:new_model}

For adding a new model, choose a model name and run
\begin{lstlisting}
AddModel["<modelname>"];
\end{lstlisting}
A sub-directory \lstinline$<modelname>$ is then created in \lstinline$"<path>/models"$, including new model, parameter and setting files. For default usage, fill out the template functions \lstinline$sv[s_]:= ... ;$ $[\sigma v_{\text{lab}}(s)]$ and \lstinline$gam[x_]:= ... ;$ $[\gamma(x)]$ in the model file, specify the DM mass \lstinline$mDM= ... ;$ ($m_{\chi}$) and internal degrees of freedom \lstinline$gDM= ... ;$ ($g_{\chi}$) in the parameter file, and save the changes.
After these four necessary specifications, the new model can already be used as described in Section~\ref{app:quick}, or  with the template script as in Section~\ref{app:tempscript}. Note that the setting file can require some accuracy control parameter adaption for optimal usage, see Section~\ref{app:settings} for rather detailed checks. For optional  usage (e.g., beyond Fokker-Planck approximation), uncomment the template routines in the model file, specify model dependent parts at the highlighted places, and change Boolean variables in the settings file according to Section~\ref{app:settings}.
 
To create the new model 
from a terminal execute
\begin{lstlisting}
> wolframscript AddModel.wls <modelname>
\end{lstlisting}

\subsection{Settings}
\label{app:settings}

Code settings are stored in \lstinline$<settings>$ files for all pre-implemented models.
These consist of accuracy control parameters and Boolean variables for optional code usage.
A possible way of directly checking the effect of these parameters is to plot the 
output variables of the relevant routines, as listed in Table~\ref{tab:primaryclassroutines}.
All parameters introduced in this section are also briefly described in the \lstinline$<settings>$ files.

\paragraph{Options}

The options for preparatory and Boltzmann solver routines are
determined by a set of global Boolean variables.
In particular, the Boolean variable {\courier RelThermalAv} determines how $\sigma v_{\text{lab}}$ is averaged in the \lstinline$PrepANN$ and \lstinline$PrepANN2$ routines. If set to true, 
$\sigma v_{\text{lab}}$ must be provided as a function of the Mandelstam $s$ variable and averages are computed fully \emph{relativistically} 
 as in Ref.~\cite{Gondolo:1990dk,Binder:2017rgn}. 
Setting {\courier RelThermalAv} to false, $\sigma v_{\text{lab}}$ must be provided as a function of 
$v_{\text{lab}}$, and the calls \lstinline$PrepANN$ and \lstinline$PrepANN2$ compute averages in 
the highly \emph{non-relativistic} limit, as in Ref.~\cite{vandenAarssen:2012ag}. The Sommerfeld 
enhanced annihilation scenario, as described in Section~\ref{sec:sommmer} and 
implemented in \textbf{SE.wl}, is 
one example where $\sigma v_{\text{lab}}$ is a sufficiently smooth function for accurate non-relativistic 
thermal averages. Especially for \lstinline$PrepANN2$ this is of great advantage since the amount of 
numerical integrals to be performed is reduced to one~\cite{vandenAarssen:2012ag}.

The Boltzmann solvers {\courier cBE} and {\courier fBE} can take \emph{optional routines} as input.
These are summarized in Table~\ref{tab:optionalroutines}.  
\begin{table}
\begin{center}
\begin{tabular}{c||c|c} 
\toprule 
   Routine (substitutes) & Implements & Returns \\ \midrule
    {\courier GetAnnMatrixAnalytic} & $ \langle \sigma v\rangle_{\theta}$ &  $\langle \sigma v\rangle_{\theta} $ \\ 
    ({\courier PrepANNtheta}) & & (as matrix) \\ \midrule 
    {\courier GetScatMatrix}  & $\langle W \rangle_{\tilde \Omega} $ &$\hat{\mathbf{C}}_{\text{el}}$ \\ 
    ({\courier PrepSCATT} for {\courier fBE})& &[see Eq.~(\ref{eq:matrix})] \\ \midrule 
    {\courier GetSecondMomentScat}  & $\langle W \rangle_{\tilde \Omega} $& $ \langle C_{\text{el}} \rangle_2$,   \\ 
({\courier PrepSCATT} for {\courier cBE})&  & $ \langle C_{\text{el}} \rangle_2^\prime$\\ \bottomrule
\end{tabular}
\end{center}
\caption{\emph{Optional input routines} that can be provided in a \lstinline$<model>$ file. The routine output shown here is symbolic and we refer to the model file \textbf{TH.wl} for a more detailed description.
}
\label{tab:optionalroutines}
\end{table}
The first routine allows for directly implementing $\langle \sigma v \rangle_\theta$ -- especially 
useful
if the angular average can be performed analytically.  
By setting the 
Boolean variable {\courier Analyticsvtheta} in \lstinline$<settings>$ to true, {\courier fBE} then adopts this 
first optional routine, making the call to \lstinline$PrepANNtheta$ redundant. The other two routines  
implement $\langle W \rangle_{\tilde \Omega} $ and allow for $C_{\text{el}}$ usage beyond the 
Fokker-Planck approximation. This is activated in {\courier cBE} and {\courier fBE}, if {\courier FullCel} is 
set to true in  \lstinline$<settings>$. \lstinline$PrepSCATT$ calls are then redundant. For a concrete 
example
implementing all these optional routines, see \textbf{TH.wl} (threshold annihilation, as in 
Section~\ref{sec:sub_threshold}).

Lastly, if the Boolean variable {\courier KDonly} is true, annihilation is switched off, and one can 
investigate kinetic decoupling only (see, e.g., Ref.~\cite{Bringmann:2009vf}) with the {\courier cBE} and 
{\courier fBE} routines. All preparing routine calls for annihilation are then redundant. For the {\courier fBE} 
routine, this also allows to check accuracy settings, see Section~\ref{sec:test}. 

\paragraph{Accuracy control parameters} 

The accuracy of preparatory and Boltzmann solver routines is controlled by a set of global parameters. In particular, the accuracy 
parameters introduced in Section~\ref{sec:impl} translate to \db internal names as follows. The  
local maximum error, $\text{err}_i$, is stored in {\courier nBEerr}, {\courier cBEerr} and {\courier fBEerr}.
Code-internal names for
$\text{errNewton}_i^{(n+1)}$ (controlling the Newton iterations) are
{\courier cBEerrNewton} and {\courier fBEerrNewton}. The maximum relative change for any Newton 
iteration depth ($n$), $\text{errMaxNewton}_i$, is set by {\courier cBEerrMaxNewton} and 
{\courier fBEerrMaxNewton}. The default values of these variables are generally not expected to require adjustments. 

The number of phase-space density elements used by the {\courier fBE} routine is set by the integer 
variable {\courier qN}. For dimensionless momenta $q=p/\sqrt{m_{\chi} T}$ in the default interval between
{\courier qmin} $=10^{-8}$ and {\courier qmax} $=12$, {\courier qN} ranges between 
about 120 and 400 in the pre-implemented models. Note that the runtime of {\courier fBE} scales quadratically with {\courier qN}, implying that it is worth 
optimizing the choice of this parameter
in a newly added model.
This can be done by, e.g., checking for numerical artifacts of too low phase-space density resolution, as described in Section~\ref{sec:test}.

We turn now to the accuracy parameters for the preparatory routines. {\courier PrepANN} and 
{\courier PrepANN2} compute $\langle \sigma v \rangle$ and $\langle \sigma v \rangle_2$, respectively. 
To do so, these routines adaptively tabulate $(\ln x , \ln \left[ \langle \sigma v \rangle_{(2)}/ \text{GeV}^{-2} \right] )$ 
pairs in a pre-set time ($x$) interval.
The initial, minimum number of table entries 
in this time interval is governed by the parameter {\courier Nx}. The 
adaptive procedure for increasing the tabulation density 
is based on the comparison of middle values, obtained from first and third order interpolations of each two 
consecutively tabulated points.
The variable {\courier iacc} and  {\courier imax} control this adaptive tabulation by setting the maximum allowed relative 
difference  between those middle values and the maximal number of refinement iterations allowed, respectively. 
{\courier pg} sets the precision goal (in digits) for the intrinsic \Mma function {\courier NIntegrate} that is used for numerical integration.  
After this construction, the tables are first interpolated and then exponentiated, resulting in the
functions {\courier svx[x]} and {\courier sv2x[x]} listed in
Table~\ref{tab:primaryclassroutines}.

{\courier PrepANNtheta} computes $\langle \sigma v \rangle_{\theta}$ by first constructing $\ln \left[ \langle \sigma v \rangle_{\theta}/ \text{GeV}^{-2} \right]$ values on an irregular quadratic grid with axes $\ln \left[ p / \text{GeV} \right]$ and $\ln \left[\tilde{p} /\text{GeV} \right]$ based on a \WL adaptive mesh 
procedure (used in plotting routines). The square root of the initial number of grid points is 
{\courier plotPoints}. The recursion limit for grid adaption is controlled via {\courier plotMaxR}. This 
irregular grid is then converted, by interpolation, into the regular, much denser grid {\courier tpregsvtheta} that is used in the actual calculations. The 
square root of the number of points in this regular grid is set by {\courier Np}, which can require values up to 
several thousand to maintain sharp features. The conversion time is rather short  
(compared to the overall \textbf{fBE} runtime), though scaling 
quadratically with {\courier Np}.%
\footnote{{\courier fBE} needs for every time step a $\text{{\courier qN}}^2$-sized matrix, whose entries are $\langle \sigma v \rangle_{\theta}$ as a function of $q_i$ and $q_j$. This matrix is provided by the sub-routine {\courier GetAnnMatrixTab}, which linearly interpolates the regular {\courier tpregsvtheta} in natural logarithm space. The regularity ensures fast access time, which overall saves much more computational time than needed for the conversion of the irregular to this regular table.}

\paragraph{Effective degrees of freedom}

The present release of \db uses the SM effective number of energy and entropy degrees of freedom from Drees \emph{et al.}~\cite{Drees:2015exa} in form of a table located in the source directory. 
Alternative tables (in the same format) can be used by adding a corresponding
file to the source directory, and replacing the string in \lstinline$dof="dof_Drees_etal.dat"$ in \lstinline$<settings>$ with this new file name. For other degree of freedom calculations, see the recent Ref.~\cite{Saikawa:2020swg} and references therein.
\bigskip

\section{Elastic scattering beyond Fokker-Planck approximation}
\label{app:full_cel}

The \db routines {\courier cBE} and {\courier fBE} can compute the relic abundance with the elastic scattering collision term $C_{\text{el}}$ once the model dependent quantity $\langle W \rangle_{\tilde \Omega} \equiv \frac{1}{4 \pi} \int \text{d} \tilde{\Omega} W$, with $W$ as introduced in Eq.~(\ref{eq:Wmain}), is provided. To illustrate this, we discuss concrete examples in Section~\ref{app:examples} where many of the integrals appearing in $\langle W \rangle_{\tilde \Omega}$ can be analytically performed, resulting in fairly simple expressions. For those cases, we also make a direct comparison between the resulting $C_{\text{el}}$ and its Fokker-Planck approximation. 
Finally, in Section~\ref{app:implementation} we explain how $C_{\text{el}}$ is implemented in the \db code and how to adjust it to any given $\langle W \rangle_{\tilde \Omega}$.

\subsection{Examples}
\label{app:examples}
As a first example, we consider {\it non-relativistic heat bath particles}, with 
$g_{\pm}(\omega)  \left[1\mp g_{\pm}(\tilde{\omega})\right] \simeq e^{-[m_f+k^2/(2m_f)]/T}$, and scattering amplitudes 
{\it independent of Mandelstam variable $s$}. 
In this case most of the integrals entering in $\langle W \rangle_{\tilde{\Omega}}$ can be solved analytically, 
leading to
\begin{align}
&\langle W \rangle_{\tilde{\Omega}}(p,\tilde{p},T) =\frac{e^{- m_f/T} T}{64 \pi g_{\chi}  E \tilde{E}p\tilde{p}} \times  \label{eq:nonrel}\\ & \int\limits_{|p-\tilde{p}|}^{p+\tilde{p}} \text{d}q \; |\mathcal{M}|^2_{\chi f\leftrightarrow\chi f} \exp\left(-\frac{ m_f}{2T} \left[\left(\frac{q^0}{q}\right)^2+\left(\frac{q}{2m_f}\right)^2 \right]\right). \nonumber
\end{align}
We note that this expression is valid also for relativistic DM, and for an arbitrary dependence of 
$|\mathcal{M}|^2_{\chi f\leftrightarrow\chi f} $ on $t=q_0^2-q^2$, where $q=|\tilde{\mathbf{p}}-\mathbf{p}|$ is 
the momentum transfer and $q^0=\tilde{E} - E$ the energy transfer. The remaining integral can be performed analytically for amplitudes with a simple power law dependence on the Mandelstam variable $t$. In particular, in the case of a
constant scattering amplitude, as in the sub-threshold annihilation scenario discussed in Section~\ref{sec:sub_threshold}, this can be further simplified to:
\begin{align}
\langle W \rangle_{\tilde{\Omega}} &= |\mathcal{M}|^2_{\chi f\leftrightarrow\chi f} \frac{ T^2}{64  g_{\chi}  E \tilde{E}p\tilde{p}}\sqrt{\frac{ m_f}{2\pi T}}  e^{- m_f/T} \label{eq:WTH}\\
&\times\bigg\{ e^{|E -\tilde{E}|/(2T)} \left( \text{erfc}\left[ a_- +b_-  \right] -\text{erfc}\left[a_+ + b_+\right]  \right)\nonumber\\&-e^{- |E -\tilde{E}|/(2T)} \left(  \text{erfc}\left[a_- - b_- \right] -\text{erfc}\left[a_+ -b_+ \right] \right)  \bigg\},\nonumber 
\end{align}
with $\text{erfc}(x)$ being the complementary error function and
\begin{align}
a_\pm&\equiv\sqrt{\frac{m_f(E -\tilde{E})^2}{2T(p\pm\tilde{p})^2}}\,,\quad b_\pm\equiv\sqrt{\frac{(p\pm\tilde{p})^2}{8m_fT}}\,.
\end{align}

\bigskip
As our second example, we consider again a scattering amplitude independent of $s$, but this time 
for {\it ultra-relativistic heat bath particles}, with $g_{\pm}(\omega)  \left[1\mp g_{\pm}(\tilde{\omega})\right] \simeq  e^{-k/T} (1 \mp e^{-\tilde{k}/T})$.
In that case, $\langle W \rangle_{\tilde{\Omega}} $ can be reduced to two remaining integrals:
\begin{align}
\langle W \rangle_{\tilde{\Omega}} &= \frac{e^{-(E-\tilde{E})/(2 T)}}{64 \pi g_{\chi}  E \tilde E p \tilde{p} } \int_{|p-\tilde{p}|}^{p+\tilde{p}} \text{d}q \; q  |\mathcal{M}|^2_{\chi f\leftrightarrow\chi f} \label{eq:Wrel} \\
& \times \int\limits_{q^0/q}^{1} \text{d}\tau \frac{\omega^{\star}}{q \tau-q^0} g_\pm(\omega^{\star})\left[1 \mp g_\pm(\tilde{\omega}^{\star})\right]. \nonumber
\end{align}
As before, this holds also for relativistic DM, and an arbitrary $t$-dependence of the scattering amplitude.
In the above expression, the energies are given by $\omega^{\star}= (q^2-q_0^2)/[2(q \tau -q^0)]$ and 
$\tilde{\omega}^{\star} = \sqrt{(\omega^{\star})^2-2\omega^{\star} q \tau + q^2}$.

\bigskip
We now turn to explicit comparisons between $C_{\text{el}}[f_{\chi}]$ calculated in these limits and the 
Fokker-Planck approximation $C_{\text{FP}}[f_{\chi}]$. For definiteness, 
and to have a non-vanishing scattering term, we choose a classical DM phase-space distribution $f_{\chi} = e^{-E/T_{\chi}}$, but with 
$T_{\chi}$ slightly different from the heat bath temperature $T$. 
We will further always consider the product of $p^4/E^2$ with the collision terms in these comparisons, 
because it is $\int \text{d}p (p^4/E^2) C_{\text{el,FP}}$ 
that determines the scattering strength $\langle C_{\text{el}} \rangle_2$  
in Eq.~\eqref{yfinalfinal}, giving the $y(x)$ evolution in the \textbf{cBE} approach.

\begin{figure}[t]
\centering
\includegraphics[scale=0.44]{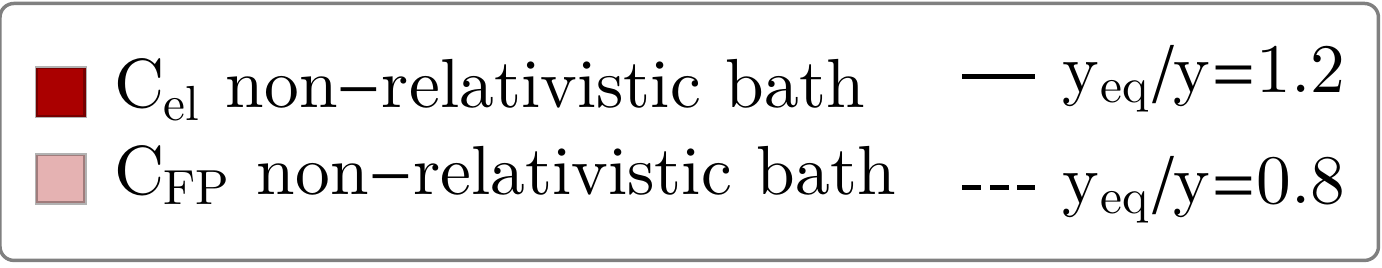}
\includegraphics[scale=0.35]{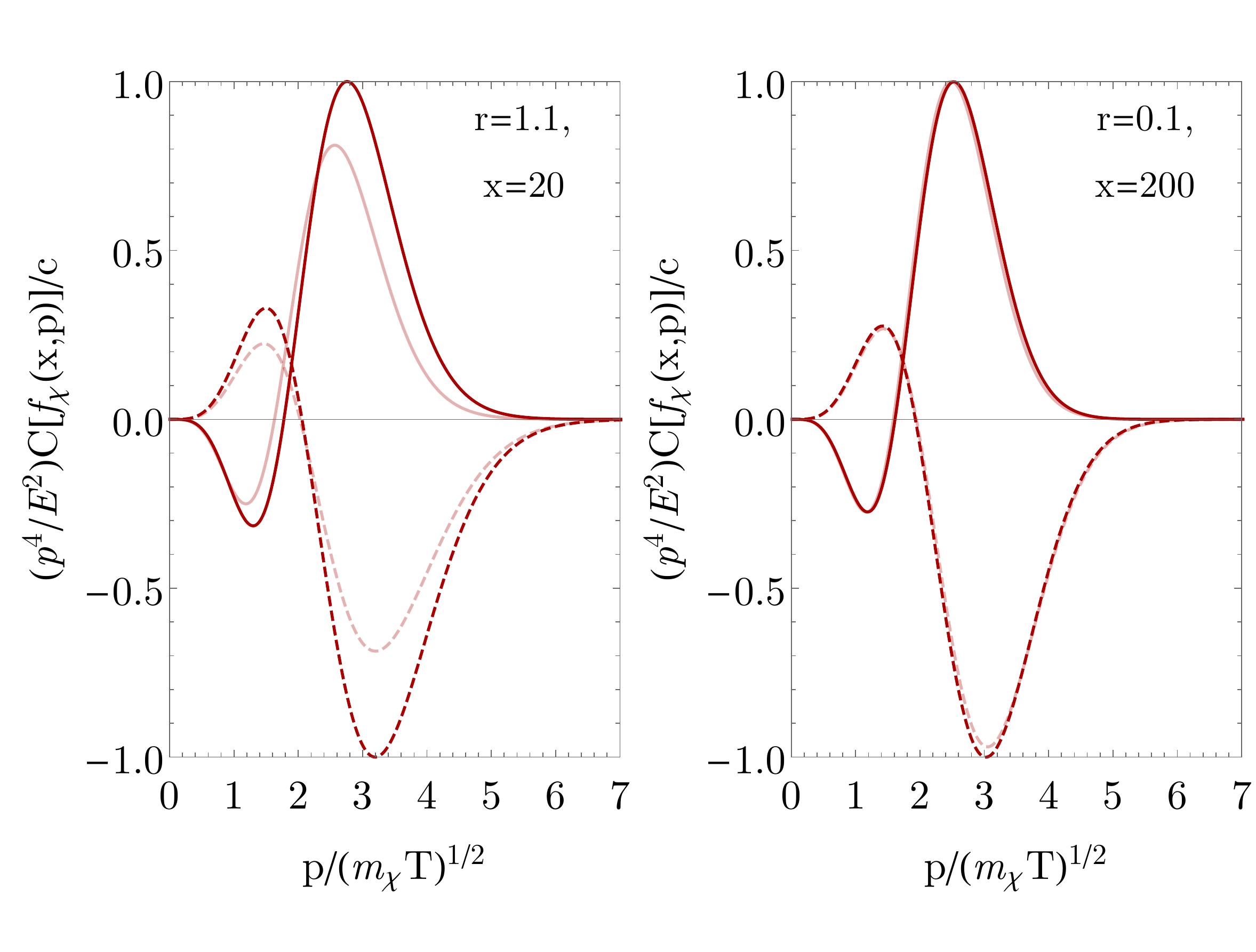}
\caption{ 
The relative amplitude of the scattering collision term,
as a function of the DM momentum $p$, for a constant scattering amplitude ($|\mathcal{M}|^2_{\chi f\leftrightarrow\chi f} = const.$) and non-relativistic bath particles. 
As indicated in the legend, the various curves correspond to different versions of the collision terms,
$C_\mathrm{FP}$ or $C_\mathrm{el}$ based on  Eq.~(\ref{eq:WTH}), as well as different temperature 
ratios $y_{\rm eq}/y=T/T_\chi$.  
Each curve is normalized to the same constant {$c= \text{Max}\left[p^4/E^2 |C_\mathrm{el,FP}|\right]$}, 
where the maximum is taken over all lines of the respective plot. 
}
\label{fig:NRCel}
\end{figure}

\begin{figure}[t]
\centering
\includegraphics[scale=0.41]{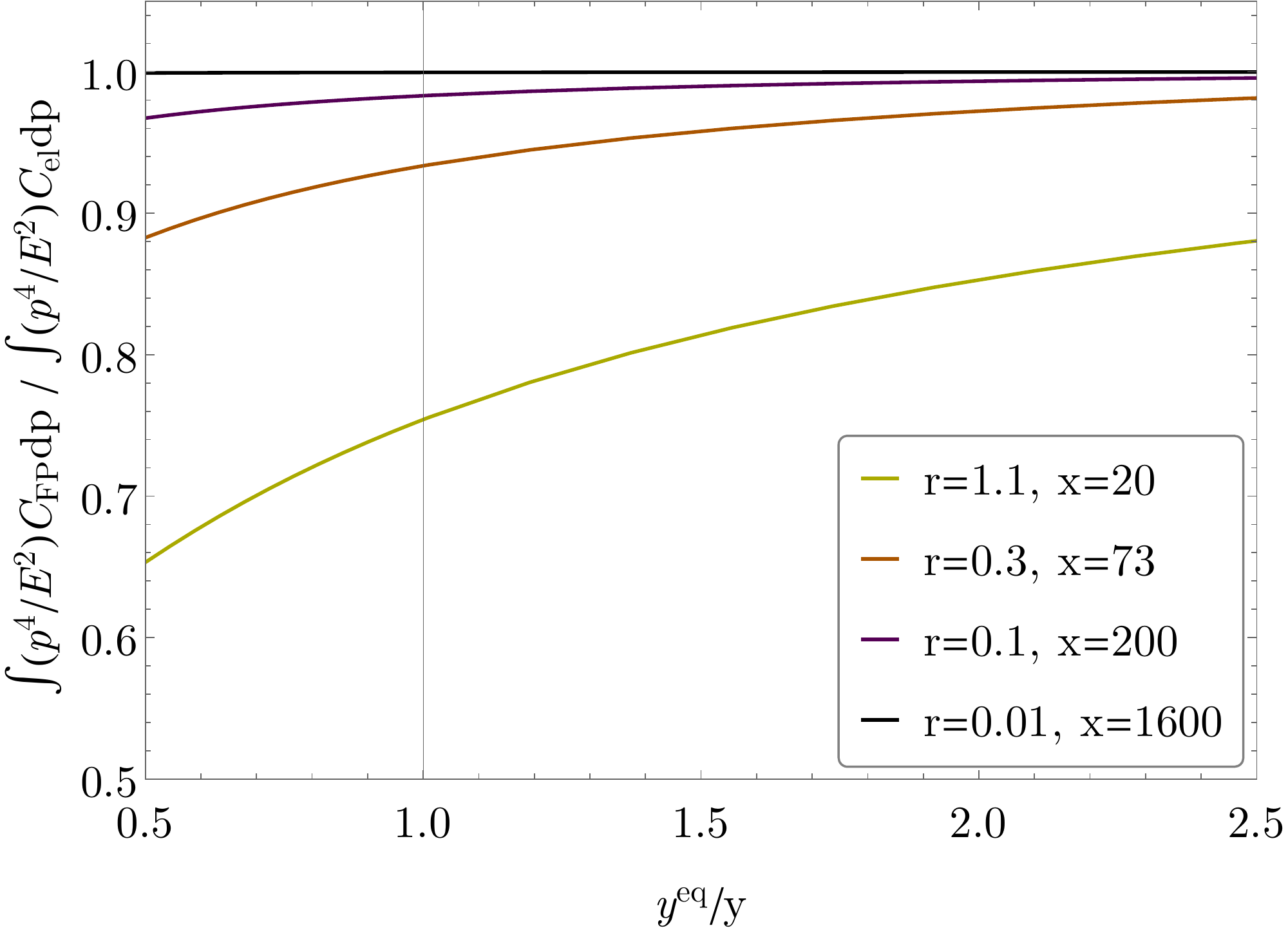}
\caption{
The magnitude of the second moment of the Fokker-Plank approximation relative to the full scattering collision term, as a function of the DM 
temperature,  for a constant scattering amplitude and non-relativistic bath particles. {We note that 
for thermally produced DM $y^{\text{eq}}/y <1$ can only be achieved
through (annihilation) processes that lead to DM self-heating.}}
\label{fig:NR2mom}
\end{figure}

\medskip
We start by comparing, in Fig.~\ref{fig:NRCel}, 
 $C_{\text{el}}$ with $\langle W \rangle_{\tilde \Omega}$ 
from Eq.~(\ref{eq:WTH}) to $C_{\text{FP}}$ with $\gamma(x)$ from Eq.~(\ref{eq:thFP}).
The particular combinations of $r$ ($=m_f/m_\chi$) and $x$ values that we use here, for illustration, are chosen such that 
{\it i)} $rx \gg 1$ to make sense of the non-relativistic bath assumption, and 
{\it ii)} $\gamma/H \sim 1$ (close to kinetic decoupling) for $|\mathcal{M}|^2_{\chi f\leftrightarrow\chi f}= \lambda^2$ with $\lambda=0.05$.
At late times, i.e.~for large $x$ values and hence $r \ll 1$, both scattering collision term descriptions are 
very close to each other (right panel). 
This is expected because small momentum transfer is guaranteed in this regime. For $r\gtrsim 1$ on the other hand, 
relevant for sub-threshold annihilations, small momentum transfer is not guaranteed. 
For example, for $r=1.1$ (left panel) the Fokker-Planck approximation underestimates the 
collision term up to 
about  30 \% at x=20 compared to the more 
accurate $C_{\text{el}}$ description (at these particular values of $y_{eq}/y$). 
Similar differences are shown in Fig.~\ref{fig:NR2mom}, 
where instead the  integrated quantities $\int \text{d}p (p^4/E^2) C_{\text{el,FP}}$, as entering \textbf{cBE}, are compared. 
This qualitatively explains the differences in the 
relic-abundance results presented in Section~\ref{sec:sub_threshold}: 
$C_{\text{el}}$ is more efficient compared to $C_{\text{FP}}$ in keeping 
DM in kinetic equilibrium, 
thus leading to a relic density closer to that of the \textbf{nBE} approach.

Finally, we compare the scattering collision terms for ultra-relativistic bath particles ($r=0$).
For illustration, we numerically integrate Eq.~(\ref{eq:Wrel}) both for a constant and a Mandelstam 
$t$-dominated amplitude, $|\mathcal{M}|^2_{\chi f\leftrightarrow\chi f}=-t/\Lambda^2$, respectively,
also for different quantum statistics of the heat bath particles.\footnote{%
For Maxwell-Boltzmann statistics, $g(\omega)=e^{-\omega/T}$, Eq.~(\ref{eq:Wrel}) simplifies 
without further approximations 
to
\begin{align}
\langle W \rangle_{\tilde{\Omega}} =  \frac{T^2}{32 \pi g_{\chi}  E \tilde E p \tilde{p}} \times \left[ c_1 e^{-|p- \tilde{p} |/(2T)} -  c_2e^{-(p+\tilde{p})/(2T)} \right],\nonumber 
\end{align}
with coefficients $c_1=c_2=|\mathcal{M}|^2_{\chi f\leftrightarrow\chi f}$ for a constant scattering amplitude, 
and $c_1=[(p- \tilde{p})^2+ 4|p- \tilde{p} |T-q_0^2+8T^2]/\Lambda^2$
and $c_2=[(p+ \tilde{p})^2+ 4(p+ \tilde{p} )T-q_0^2+8T^2]/\Lambda^2$ for 
$|\mathcal{M}|^2_{\chi f\leftrightarrow\chi f}=-t/\Lambda^2$. One can notice that corrections coming from typical 
DM momenta, $(p+\tilde{p})\sim \sqrt{m_{\chi} T}$, are exponentially suppressed for non-relativistic dark matter as 
$\sim e^{-\sqrt{x}}$. Due to this suppression, indeed, the much simpler $C_\text{FP}$ approximates the $C_\text{el}$ 
collision term in an acceptable way even for $x\sim20$.
} 
\begin{figure}[t]
\centering
\includegraphics[scale=0.45]{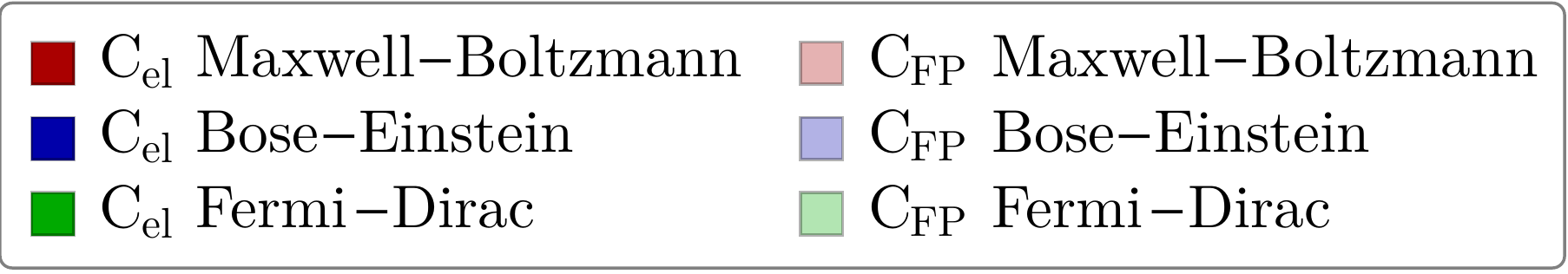}
\includegraphics[scale=0.35]{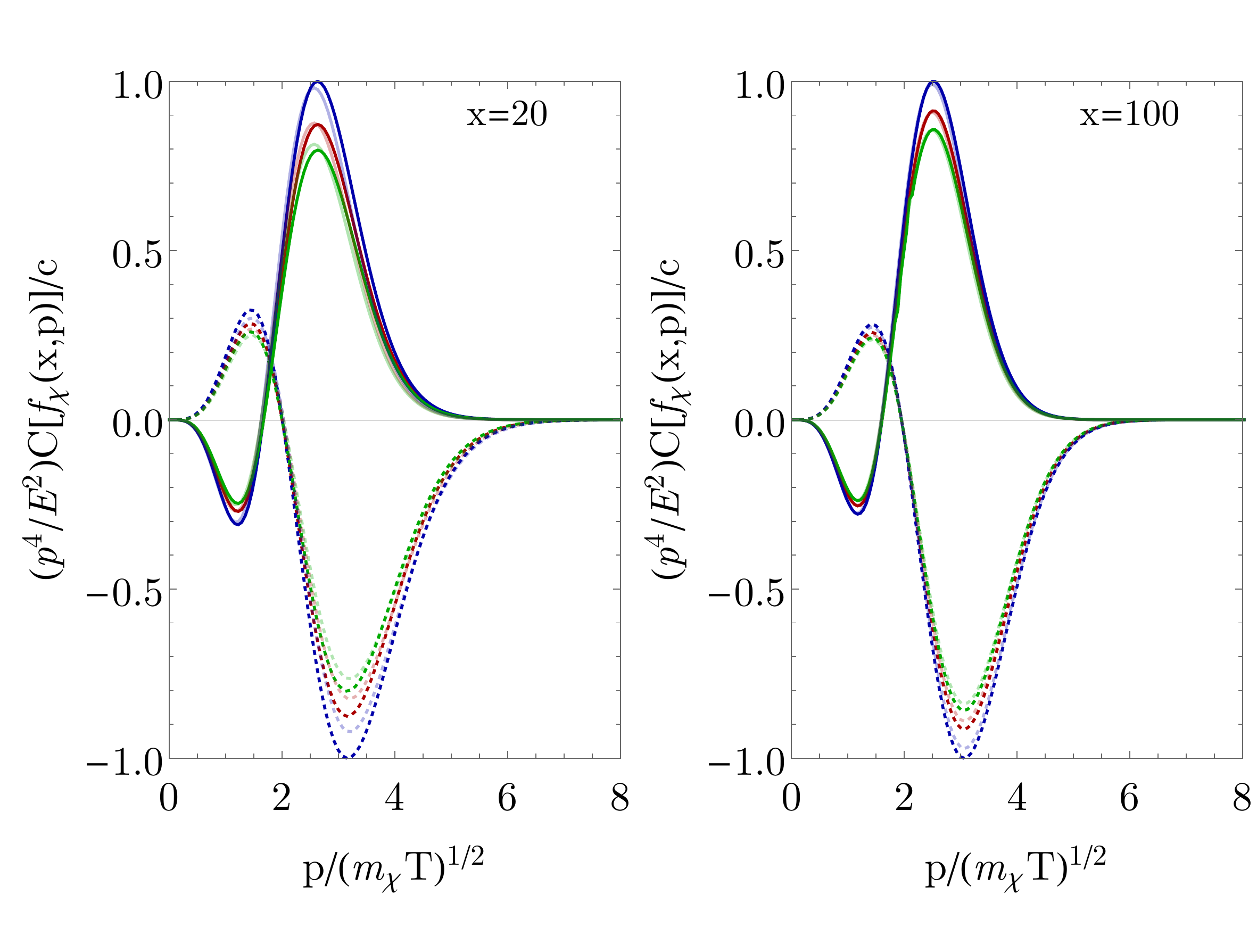}
\caption{Same as in Fig.~\ref{fig:NRCel}, but for the case of ultra-relativistic bath particles with $r=0$ 
(and constant scattering amplitude).}
\label{fig:cCel}
\end{figure}
\begin{figure}[t]
\centering
\includegraphics[scale=0.35]{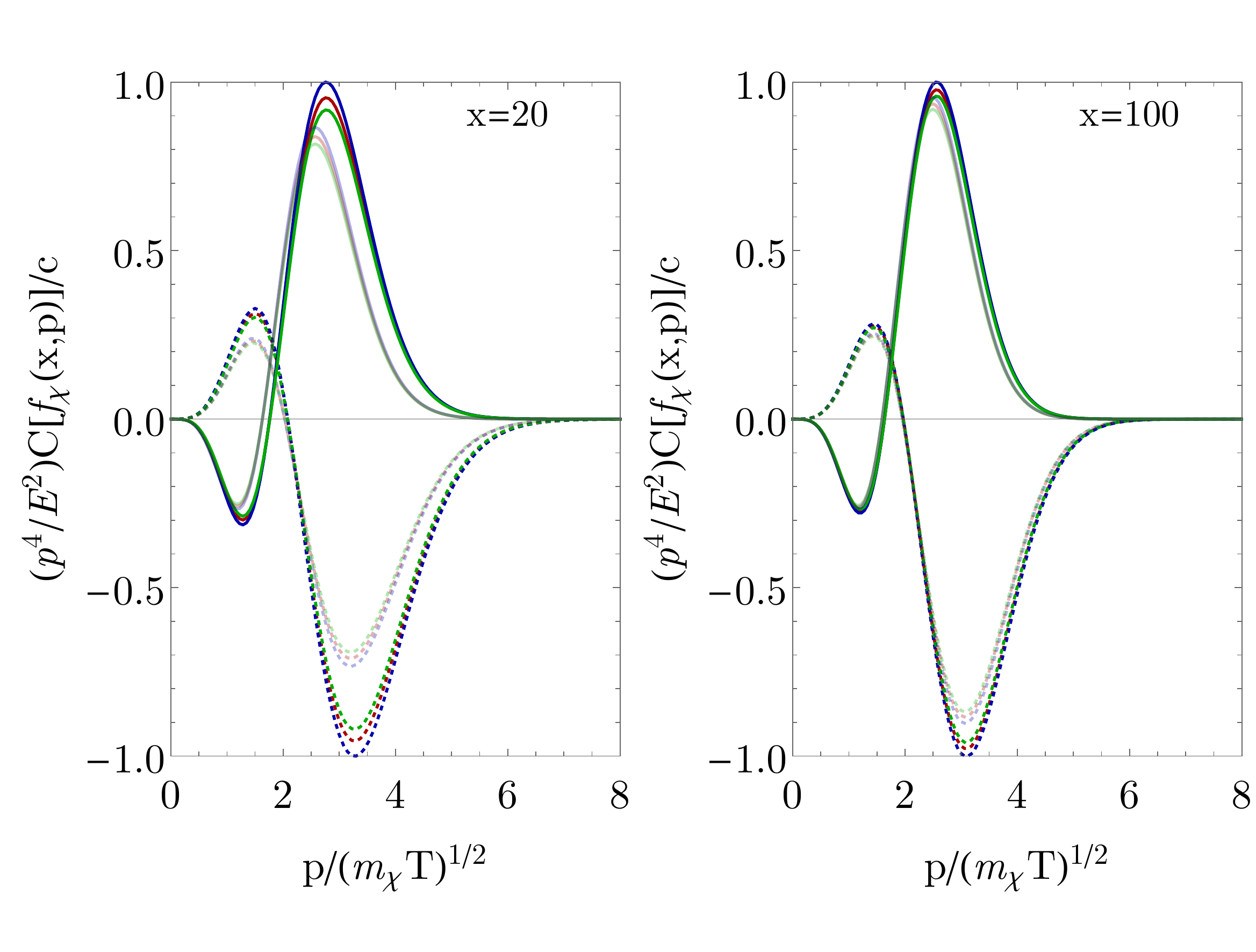}
\caption{Same as in Fig.~\ref{fig:cCel}, but for a scattering amplitude proportional to Mandelstam $t$.}
\label{fig:tCel}
\end{figure}
These results are compared to the corresponding Fokker-Planck approximation in 
Fig.~\ref{fig:cCel}
and Fig.~\ref{fig:tCel}, respectively. For low temperatures (right panels), as expected, the Fokker-Planck scattering term 
is always an excellent approximation. In the high-temperature regime (left panels), on the other hand,
less separated scales 
(the momentum transfer scale $\sim T$ versus typical DM momenta
$\sim \sqrt{m_{\chi} T_{\chi}}$) can become an issue for the Fokker-Planck approximation
--- especially if amplitudes support larger momentum transfers as, e.g., in the 
$|\mathcal{M}|^2_{\chi f\leftrightarrow\chi f}=-t/\Lambda^2$ example. 

We conclude that the error induced when adopting the much simpler Fokker-Planck approximation, 
at temperatures relevant for chemical freeze-out, depends on the exact form of the scattering amplitude.
The potential impact on the relic density in scenarios with early kinetic decoupling
is therefore also unavoidably model-dependent. However, as for example in the 
threshold scenario discussed in Section~\ref{sec:sub_threshold}, we expect that the {\it dominant} effect of 
early kinetic decoupling on the relic density
can in many cases still be fairly 
well captured by the Fokker-Planck approximation.

\subsection{Implementation}
\label{app:implementation}

In discretized form, the full elastic collision term can be written as a linear operator of the form $C_{\text{el}}[f_{\chi}]/E=
\hat{\mathbf{C}}_{\text{el}}\cdot\mathbf{f}$
with $\mathbf{f}=(f_0,...,f_N)^T$ being the momentum-discretized phase-space density $f_{\chi}$. The matrix 
$\hat{\mathbf{C}}_{\text{el}}$ can be written on a momentum grid, with $p,\tilde{p} \in \{p_0,...,p_N\} $ and 
equal spacing $\Delta p$, as:
\begin{align}
&\hat{\mathbf{C}}_{\text{el}} = -\frac{\Delta p}{2 \pi^2} \label{eq:matrix}\times\\
&
\begin{pmatrix}
\sum\limits_{j \neq 0}^{N} \langle W \rangle_{0j} p_j^2 e^{\beta(E_0-E_j)/2} & -\langle W \rangle_{01} p_1^2 e^{-\beta(E_0-E_1)/2} & ...\\
-\langle W \rangle_{10} {p_0}^2 e^{-\beta(E_1-E_0)/2} & \sum\limits_{j \neq 1}^{N} \langle W \rangle_{1j} p_j^2 e^{\beta(E_1-E_j)/2} & ...\\
... & ... & ...
\end{pmatrix},
\nonumber
\end{align}
where $\beta \equiv 1/T$ and we introduced $\langle W \rangle_{ij}\equiv \langle W \rangle_{\tilde \Omega}(p_i,p_j,T)$ 
on the momentum grid. The routines
returning the matrix in Eq.~(\ref{eq:matrix}) needed for \textbf{fBE}, and its descretized second momentum moment 
needed for the \textbf{cBE}, have the  internal \db names {\courier GetScatMatrix} and 
{\courier GetSecondMomentScat}, respectively (see also Table~\ref{tab:optionalroutines}). 

For the sub-threshold annihilation case, these two routines can be found in the 
model file \textbf{TH.wl}, where Eq.~(\ref{eq:matrix}) with $\langle W \rangle_{\tilde \Omega}$ as in 
Eq.~(\ref{eq:WTH}) is implemented. To switch from the Fokker-Planck approximation to full $C_{\text{el}}$ in 
the \textbf{cBE} and \textbf{fBE} approaches, set \lstinline$FullCel=True$ in the setting file.
These two routines can be adopted to other scattering scenarios, by replacing the $\langle W \rangle_{ij}$ dependent parts inside the routines.

\bibliography{Boltzmann}

\end{document}